\documentclass[a4paper,fleqn,usenatbib]{mnras}

\usepackage{graphicx}
\usepackage{mathptmx}
\usepackage{savesym}
\usepackage{amsmath}
\usepackage{txfonts}
\usepackage[T1]{fontenc}
\usepackage{ae,aecompl}

\usepackage{amssymb}	% Extra maths symbols
\usepackage{gensymb}
\usepackage{longtable,lscape}
\title[The open cluster Berkeley~90]{The little-studied cluster Berkeley~90. III. Cluster parameters}

\author[A. Marco and I. Negueruela]{
Amparo Marco$^{1,2}$\thanks{E-mail: amparo.marco@ua.es (AM)} and Ignacio Negueruela$^{1}$\thanks{E-mail: ignacio.negueruela@ua.es (IN)}
\\
$^{1}$DFISTS, EPS, Universidad de Alicante, Carretera San Vicente del Raspeig s/n,  E03690, San Vicente del Raspeig, Spain\\
$^{2}$Department of Astronomy, University of Florida, 211 Bryant Space Science Center, Gainesville, FL 32611, USA\\
}

\date{Accepted XXX. Received YYY; in original form ZZZ}

\pubyear{2016}

\begin{document}
\label{firstpage}
\pagerange{\pageref{firstpage}--\pageref{lastpage}}
\maketitle

\begin{abstract}

The open cluster Berkeley~90 is the home to one of the most massive binary systems in the Galaxy, LS\,III $+46\degr$11, formed by two identical, very massive stars (O3.5\,If* + O3.5\,If*), and a second early-O system (LS\,III $+46\degr$12 with an O4.5\,IV((f)) component at least). Stars with spectral types earlier than O4 are very scarce in the Milky Way, with no more than 20 examples. The formation of such massive stars is still an open question today, and thus the study of the environments where the most massive stars are found can shed some light on this topic. To this aim, we determine the properties and characterize the population of Berkeley~90 using optical, near-infrared and \textit{WISE} photometry and optical spectroscopy. This is the first determination of these parameters with accuracy. We find a distance of $3.5^{+0.5}_{-0.5}$ kpc and a maximum age of 3~Ma. The cluster mass is around $1\,000\:$M$_{\sun}$ (perhaps reaching $1\,500\:$M$_{\sun}$ if the surrounding population is added), and we do not detect candidate runaway stars in the area. There is a second population of young stars to the Southeast of the cluster that may have formed at the same time or slightly later, with some evidence for low-activity ongoing star formation.      
%This is a simple template for authors to write new MNRAS papers.
%The abstract should briefly describe the aims, methods, and main results of the paper.
%It should be a single paragraph not more than 250 words (200 words for Letters).
%No references should appear in the abstract.
\end{abstract}


%%%%%%%%%%%%%%%%%%%%%%%%%%%%%%%%%%%%%%%%%%%%%%%%%%

%%%%%%%%%%%%%%%%% BODY OF PAPER %%%%%%%%%%%%%%%%%%

\section{Introduction}

How do massive stars form? A number of different hypotheses have been put forth along time to try to circumvent the obvious physical difficulties that the formation of a very massive star presents \citep[for a review]{zy07}. At present, two main theories are invoked: (a) monolithic core accretion, basically a scaled-up version of classical low-mass formation theories, where very high opacities allow infalling material to overcome the radiation pressure \citep[e.g.][]{ys02,krumholz2009} and (b) competitive accretion, where massive stars are formed in cluster cores, benefiting from the gravitational potential of the whole cluster to accrete more material \citep[e.g.][]{bonbate06,smith2009}.

The validity of these theories must be supported by the results of observations. The idea of competitive accretion stems naturally from the widespread evidence that massive stars are found mostly in young clusters. Because of their rarity, simple statistical arguments would suggest that they should preferentially be found in the most massive clusters \citep{elmegreen1983}. Indeed, observations of clusters show that, in general, the mass of the most massive cluster member is a function of the cluster mass \citep{larson1982,elmegreen1983,weidner2010}. However, it is still unclear if this is simply a consequence of statistical sampling of the initial mass function (IMF) or a consequence of a deeper underlying principle \citep[resulting in sorted sampling; ][]{weidner2013}. Moreover, there is direct evidence of discs around some moderately massive stars \citep[up to $M_{*}\sim25$--30$\:$M$_{\sun}$; see][for a review]{beltran16}, supporting the idea of monolithic collapse. On the other hand, massive stars in young clusters seem to lie in the regions of highest stellar density \citep{rivilla2013,rivilla2014}, a key prediction of the competitive accretion scenario that seems at odds with monolithic collapse.

One of the key observations that may help clarify theories of massive star formation is the degree of isolation under which massive stars can be formed \citep{lamb2010}. \citet{oey2013} presented a sample of 14 OB stars in the Small Magellanic Cloud (SMC) which meet strong criteria for having formed under extremely sparse star-forming conditions in the field. As this study was performed in the SMC (at a distance modulus of almost 19~mag), the stars observed can be safely claimed to be free of bright neighbours, but may still have a substantial complement of low-mass stars. In the Milky Way, similar observations can reach fainter magnitudes and thus imply certainty of isolation \citep[e.g.][]{dewit2004}. Characterization of the environment of very massive stars is a powerful tool not only to investigate their formation process, but also to understand how the IMF is sampled.

%Much evidence indicates that stars prefer to form in groups of various sizes rather than
%singly. On the smallest scales, they belong to binary or multiple system and on larger scales they are formed as members of larger groups or clusters \citep{larson1995}. 

In this context, this is the third paper in a series of publications relating to the open cluster Berkeley~90 to study different astrophysical issues.
\citet[from now on, Paper~I]{jesus2015a} characterized the two brightest stars in the cluster: LS\,III $+46\degr$11 and LS\,III $+46\degr$12. The former was identified as a new very early O-type spectroscopic binary (O3.5\,If* + O3.5\,If*), while the latter was classified as O4.5\,V((f)) and suspected of binarity.  In addition, orbital parameters for LS\,III $+46\degr$11 were derived by using multiepoch optical spectra, finding that it is composed by two very similar stars in a highly eccentric 97.3~d orbit. Moreover, LS\,III $+46\degr$11 is one of only six objects in the Northern hemisphere with spectral type earlier than O4 \citep{jesus2016}. The luminosity classification of LS\,III $+46\degr$12 was changed from V to IV in \citet{jesus2016}, with the introduction of this new luminosity subclass. In Paper II \citep{jesus2015b}, the ISM sightlines towards LS\,III $+46\degr$11, LS\,III $+46\degr$12 and another B-type member (there named LS\,III $+46\degr$11B) were studied, with the conclusion that the absorbing ISM has at least two clouds at different velocities: one with a lower column density (thinner) in the \ion{K}{i} lines located away from Berkeley~90 with velocity $\approx-7\:$km$\,\textrm{s}^{-1}$, and a second one with a higher column density (thicker), perhaps close to the cluster, with velocity $\approx-20\:$km$\,\textrm{s}^{-1}$. 

\defcitealias{jesus2015a}{Paper~I}

There is very little previous information about the open cluster Berkeley~90 itself.
As a result of a systematic inspection of the Palomar Sky Survey plates, \citet{setteducati1962} identified 91 new open clusters that in the subsequent literature were named Berkeley clusters, where the majority of the stars are fainter than 13th magnitude. Berkeley~90 is an open cluster located within the isolated \ion{H}{ii} region Sh2-115 ($l=84\fdg9$), away from the Galactic plane ($b=+3\fdg94$). \citet{sanduleak1974} was the first author to notice the presence of two OB stars (LS\,III $+46\degr$11 and LS\,III $+46\degr$12) in Berkeley~90. The first study of the cluster using 2MASS photometry and proper motions from NOMAD (Naval Observatory Merged Astrometric Dataset) was conducted by \citet{tadross2008}, who obtained a new centre of coordinates ($\alpha$(J2000) = 20h 35m 17s, $\delta$(J2000)=$+46\degree$~$50\arcmin$~$06\arcsec$), and estimated a diameter of $5\arcmin$, an age of 0.1~Ga (clearly too old for a cluster containing OB stars), a reddening of $E(B-V)=1.15$~mag, and a distance of $2430\pm70$~pc. 

In this paper, we present optical and near-infrared (near-IR) photometry for Be~90, and optical spectroscopy for most of the cluster members in the upper main sequence. With this dataset, we are able to determine the parameters of the cluster with better accuracy and estimate the mass of the cluster, thus characterizing the  properties and population of a small cluster hosting one of the most massive systems in the Galaxy. Furthermore, to understand the cluster formation process, we investigate its surroundings. For this purpose, we carry out the same study for a region to the southwest of the cluster, where bright-rimmed clouds are present. This area contains substantial amounts of gas and dust.

The paper is divided following this structure: in Section~\ref{data}, we present the observations and reduction procedures used; in Section~\ref{results} we determine the parameters of the cluster and characterize its population, and in Section~\ref{discussion} we calculate the mass of the cluster, look for the existence of possible runaway stars and study the environment of other O3-type stars in the Galaxy. Finally, we list the conclusions.

\section{Observations and data}
\label{data}

We used the imager and spectrograph Andaluc\'{i}a Faint Object Spectrograph and Camera (ALFOSC) on the Nordic Optical Telescope (NOT) at the La Palma observatory to obtain $UBVR$ photometry on the night of 2007 July 9, and spectroscopy of 20 selected stars on the nights of 2004 October 4, 2005 October 2\,--\,4, and 2007 July 10.
We also downloaded UKIDSS\footnote{https://www.ukidss.org/archive/archive.html} \citep{lawrence2007} images in the $JHK_{{\textrm S}}$ filters to perform deep near-IR photometry.  

\subsection{Spectroscopy}

In spectroscopic mode, we used grisms \#14 and \#16 combined
with  a  range of slit widths adapted to the weather conditions (between 1$\arcsec$ and 1.8$\arcsec$) to  obtain  intermediate  resolution  spectroscopy in the classification region. Grism \#14 covers the 3275\,--\,6125\,\AA\, range with a nominal dispersion 1.4\,\AA/pixel. Grism \#16 covers the 3500\,--\,5060\,\AA\, range with
a nominal dispersion of 0.8\,\AA/pixel. The reduction procedure is the same as used in section 2.2 of \citet{marco2011}.
We observed selected bright stars in the zone and a number of objects in the central concentration whose 2MASS colours suggested membership applying criteria from \citet{negueruela2007}. We can see in Table~\ref{tab:spTBe90M} the names of the 20 observed stars (+2 stars from the literature) and their date of observation. 

For one star (later named star \textbf{238}) that presented photometric characteristics typical of an early-type emission line star, we obtained a moderately-high resolution spectrum in the H$\alpha$ region using the VPH grism \#17. This grism covers the 6330\,--\,6870\,{\bf \AA} range with a nominal dispersion of 0.26\,\AA/pixel. 

\begin{table*}
	\centering
	\caption{Stars with classification spectra taken with ALFOSC. Top panel are stars considered members and bottom panel are non-members stars\label{tab:spTBe90M}}
	\begin{tabular}{llccc} 
	  \hline
           \noalign{\smallskip}
           RA(J2000)&DEC(J2000)&Name&Spectral type& Date of observation\\
            \noalign{\smallskip}
\hline
\hline
 \noalign{\smallskip}
20:35:08.00&	$+$46:49:32.2&	36&		O8\,V	&2005 Oct 2\\
20:35:15.79&	$+$46:49:49.8&	47&		B2\,V	&2007 Sep 10\\
20:35:14.23&	$+$46:50:11.9&	57&		B0.5\,V	&2005 Oct 3\\
20:35:20.48&	$+$46:50:13.3&	59&		B2\,V	&2007 Sep 10\\
20:35:16.30&	$+$46:50:33.5&	67&		B2\,V	&2007 Sep 10\\
20:35:13.32&	$+$46:50:38.5&	70&	        B1.5\,V	&2005 Oct 3\\
20:35:07.02&	$+$46:50:45.7&	75&      	B1.5\,V	&2005 Oct 4\\
20:34:59.87&	$+$46:50:59.1&	88&		B2\,V	&2007 Sep 10\\
20:35:15.09&	$+$46:51:05.2&	90&		B1.5\,V	&2007 Sep 10\\
20:35:10.11&	$+$46:51:06.8&	92&		B2.5\,V	&2005 Oct 3\\
20:35:18.56&	$+$46:50:02.9&	176&	        O4.5\,IVf$^{,1}$&$-$	\\
20:35:12.64&	$+$46:51:12.1&	178&	        O3.5\,If$^{*}$+O3.5\,If$^{*}$$^{,1}$&$-$\\ 
20:35:18.83&    $+$46:46:42.8&  283&            O9.5\,V	&2005 Oct 2\\
\noalign{\smallskip}
\hline
\noalign{\smallskip}
20:35:20.97&	$+$46:48:36.9&	14&	B9\,III& 2005 Oct 3\\
20:34:55.62&	$+$46:50:45.2&	74&	K1\,V& 2005 Oct 4\\
20:35:18.14&	$+$46:50:52.6&	81&	F6\,III& 2004 Oct 4\\
20:35:10.26&	$+$46:51:36.4&	109&	G2\,III& 2004 Oct 4\\
20:35:07.45&	$+$46:51:36.8&	110&	F6\,V& 2004 Oct 4\\
20:34:56.08&	$+$46:51:38.9&	111&	K0\,IV&2005 Oct 2\\
20:34:53.47&	$+$46:52:07.0&	125&	K0\,III&2005 Oct 2\\
20:35:09.55&	$+$46:52:19.9&	134&    F8\,IV&2004 Oct 4\\
20:35:22.02&	$+$46:51:51.9&	179&	B8\,III&2004 Oct 4\\
\noalign{\smallskip}
\hline
\end{tabular}
\begin{list}{}{}
\item[]$^{1}$ taken from \citet{jesus2015a}
\end{list}
\end{table*}

\subsection{Optical photometry and near-IR photometry}
\label{opt_phot}

In imaging mode, the camera covers a field of $6\farcm5 \times 6\farcm5$ and has a pixel scale of $0\farcs19\:$pixel$^{-1}$. We took the $UBVR$ photometry from two regions: one of them  centred on the nominal position of the open cluster Berkeley~90 and the other one to the southeast of this first region, with a little overlap, coming close to bright-rimmed clouds associated with Sh2-115 (see Figure~\ref{mapmembers}). For both frames, we obtained three series of different exposure times in each filter to achieve accurate photometry for a broad magnitude range. The central positions of each frame and the exposure times used are presented in Table~\ref{tab:tab1}. 

The reduction procedure for standard and target fields is the same as used in section 2.1 of \citet{marco2011}. In this run, 12 standard stars were selected in the SA~110 and SA~113 fields from the list of \citet{landolt1992} and an aperture of 20 pixels was taken for all filters . 

We obtained the $UBVR$ photometry for 314 stars. The number of stars detected in all filters is limited by the long exposure time in the $U$ filter. In Table~\ref{tab:coorBe90} we list all the stars with complete photometry, their coordinates in J2000, their values of $V$, $(B-V)$, $(U-B)$, $(V-R)$ with the corresponding errors (the standard deviation when there are more than one measurement for each value, or the addition in quadrature of the contribution to the total error made by each photometric individual error given by {\sc daophot} in the opposite case) and their number of measurements for each magnitude or colour.

\begin{table*}
	\centering
	\caption{Log of the photometric observations taken at the NOT in 2007 July. \label{tab:tab1}}
	\begin{tabular}{lll} 
		\hline
		Be~90& RA(J2000)=20h 35m 9.8s& DEC(J2000)= $+46\degr\,51\arcmin\,13\farcs9$ \\
		Southeast Be~90& RA(J2000)=20h 35m 28.0s & DEC(J2000)= $+46\degr\,46\arcmin\,52\farcs7$ \\
		\hline
	\end{tabular}
        \begin{tabular}{lccc}
        &&Exposure times (s)&\\
       Filter&Long times& Intermediate times&Short times\\
        \hline
        $U$&300&120&60\\
        $B$&90&30&4\\
        $V$&20&6&2\\
        $R$&10&4&2\\
        \hline
        \end{tabular}
\end{table*}

We took three images in the $J$, $H$ and $K_{{\textrm S}}$ filters centred on the cluster with a size of $13\farcm0 \times 13\farcm0$ from UKIDSS \citep{lawrence2007}. The date of the observation is 2011 August 2. The specifications of the camera are given in \citet{casali2007}. The total exposure time for each filter is 40~s. The procedure for obtaining the instrumental photometry and carry out the transformation from these instrumental magnitudes to the 2MASS magnitude system \citep{Skrutskie2006} is the same that we implemented in section 2.2 of \citet{marco2016}.
We obtained photometry for 2365 stars. In Table~\ref{tab:ir} we list their coordinates in J2000 and their $J$, $H$ and $K_{{\textrm S}}$ magnitudes with the errors given by {\sc daophot} (because the number of measurements is only one for each filter). Astrometric referencing of all our images was made using the same procedure as in \citet{marco2016}. 

\section{Results}
\label{results}
		  
\subsection{Spectral classification}

\begin{figure}
\resizebox{\columnwidth}{!}{\includegraphics[angle=-90]{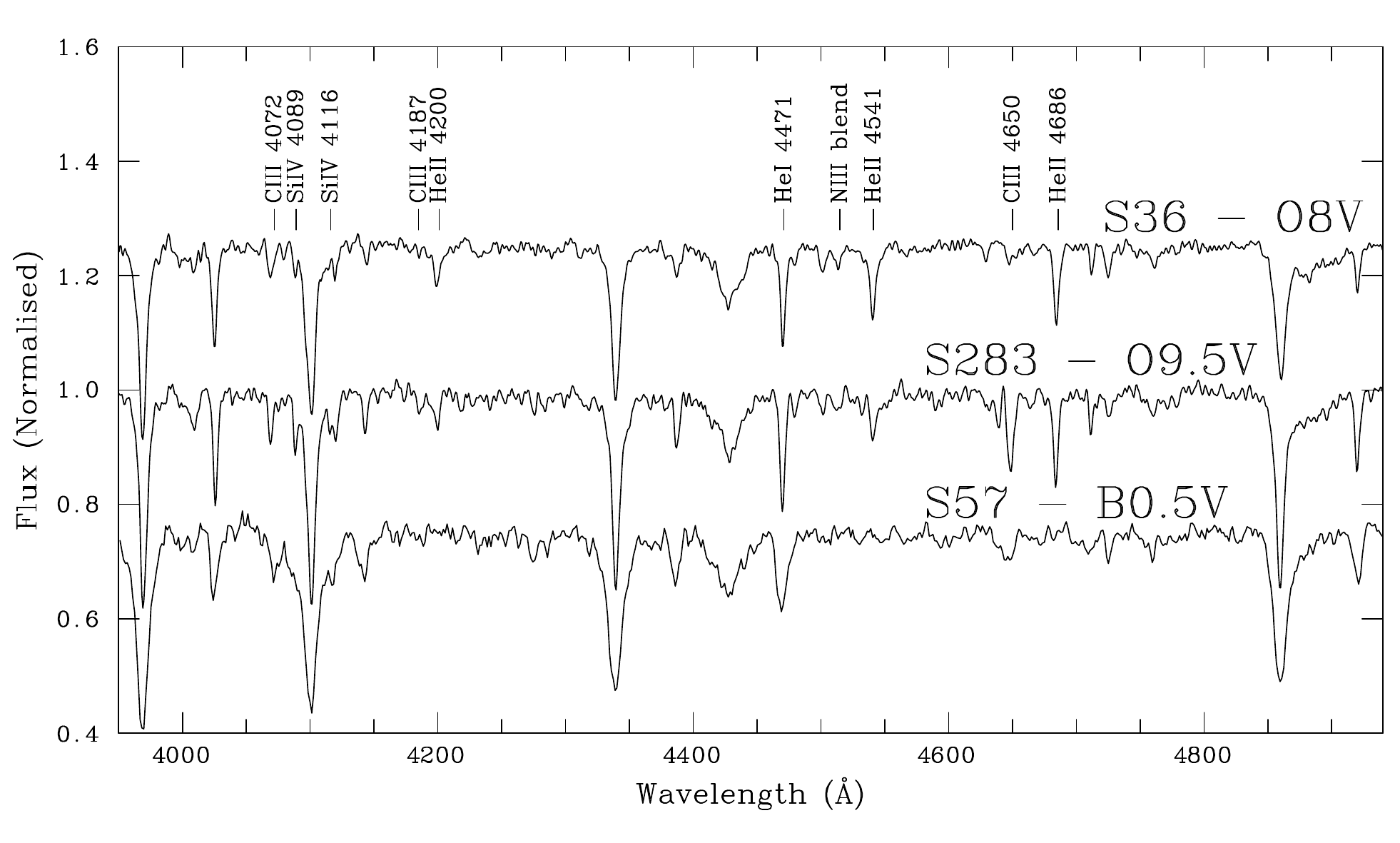}}
\centering
\caption{Spectra of the brightest members in Berkeley~90 observed with ALFOSC. Lines relevant for spectral classification are marked.\label{Amembers_spec}} 
\end{figure}

\begin{figure}
\resizebox{\columnwidth}{!}{\includegraphics[angle=0]{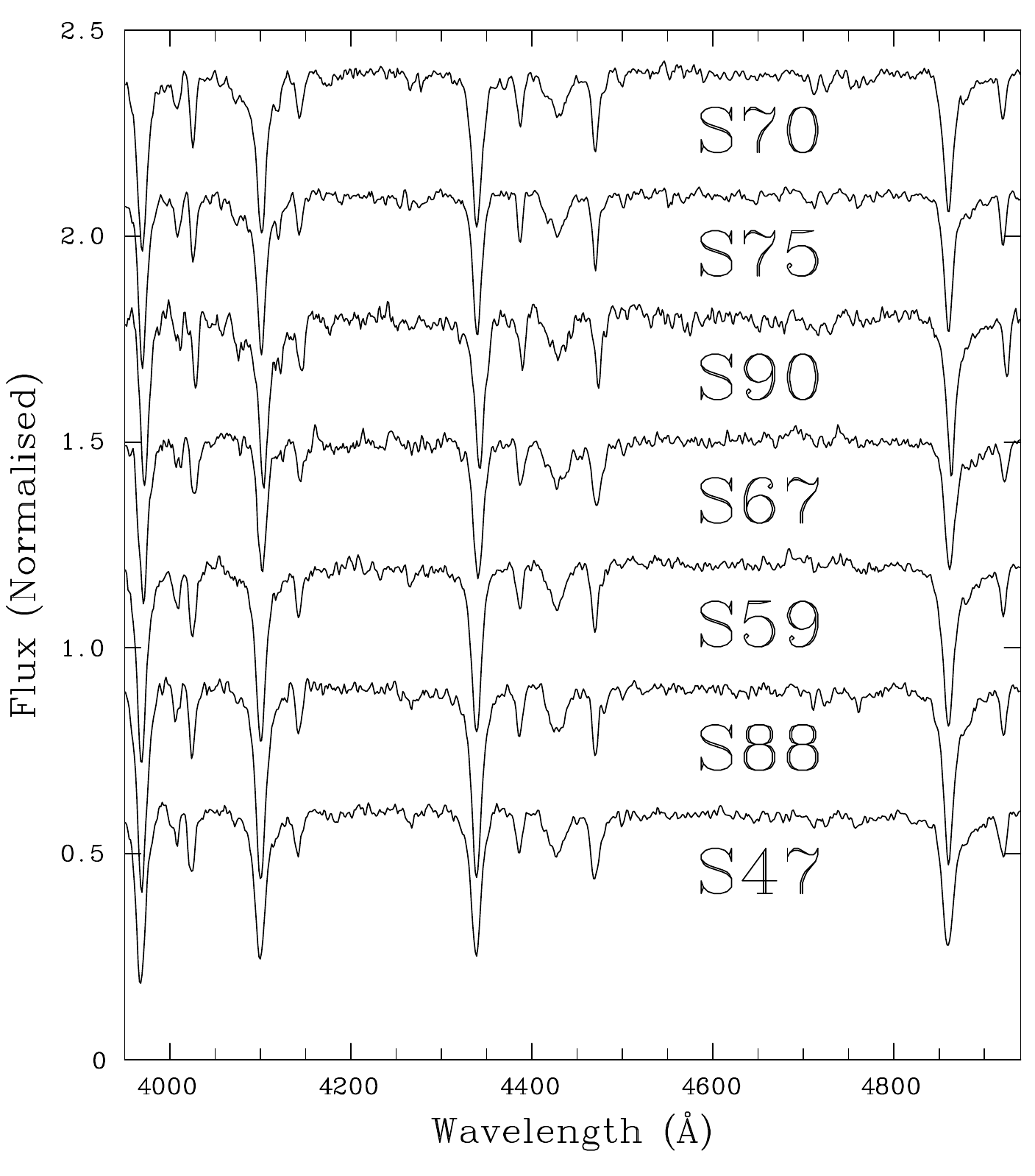}}
\centering
\caption{Spectra of other early-B members in Berkeley~90.\label{Bmembers_spec}} 
\end{figure}

We obtained spectral classifications by comparing our spectra to MK standard stars observed at a similar resolution. The comparison was carried out by eye. The standards used are those listed in \citet{negue2004}. For internal consistency, all the spectra were also compared among themselves. Most of the members observed have spectral types in the B1\,--\,B2 range. Given the moderate signal to noise, these spectral types can be considered accurate to better than one spectral subtype, in the sense that a star classified B2 is unlikely to be B1, but could be B1.5 or B2.5. For the late-type non-member stars, the classification is less accurate, as the grid of standards is not so well populated, and we consider it to be correct to $\pm$2 spectral subtypes. The spectral types for stars considered members and non-members are listed in Table~\ref{tab:spTBe90M}. In Figures~\ref{Amembers_spec} and~\ref{Bmembers_spec}, we display the classification spectra for spectroscopically observed members of Be~90. All the stars shown lie in the central concentration of Berkeley~90, except star 283 that is located in the southeast frame. The spectrum of the star 92 is not shown because of rather low signal to noise.

\subsection{HR diagrams}
\subsubsection{Optical photometry}
\begin{figure*}
\centering
\resizebox{\columnwidth}{!}{\includegraphics[clip]{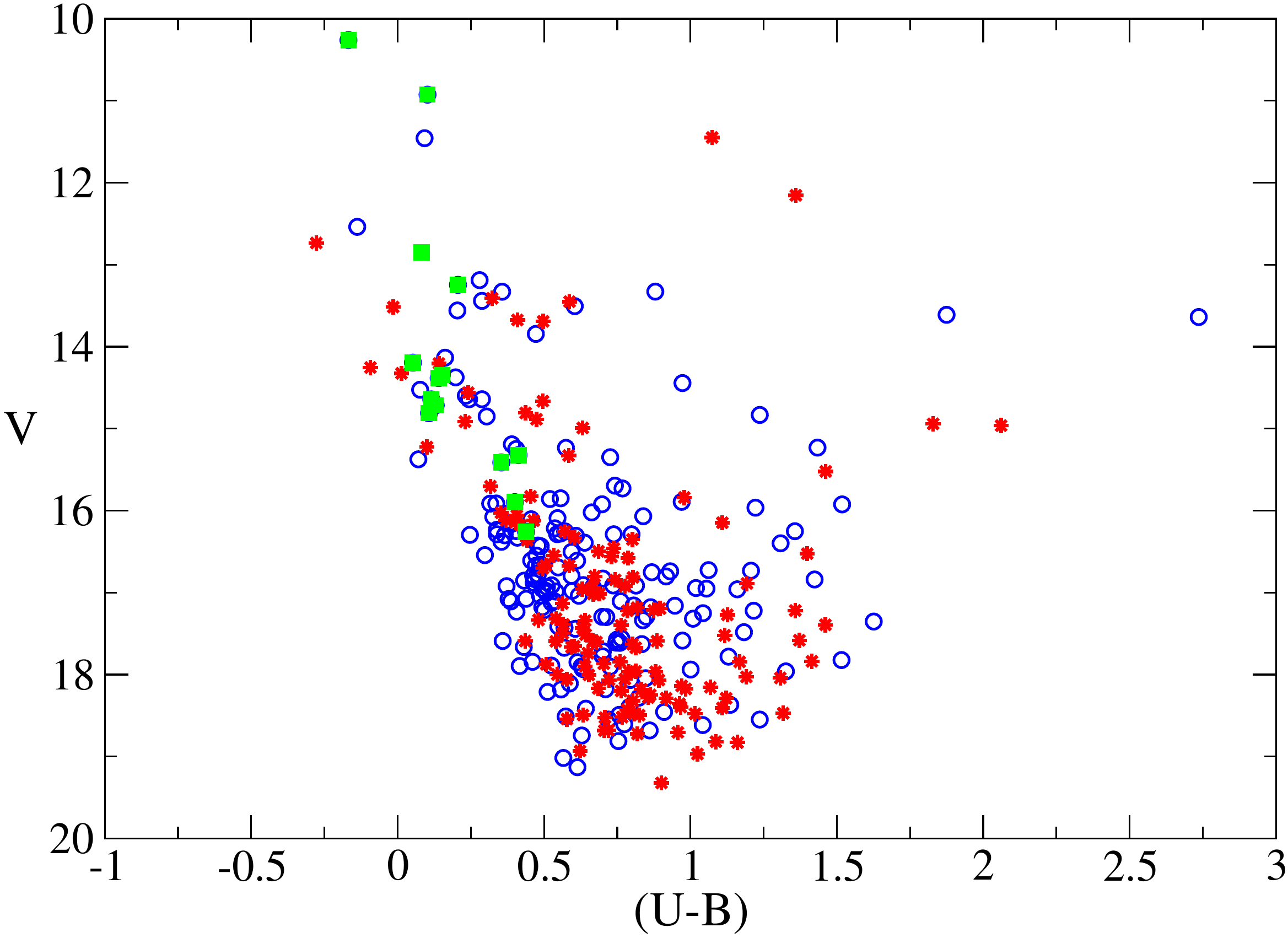}
}
\resizebox{\columnwidth}{!}{\includegraphics[clip]{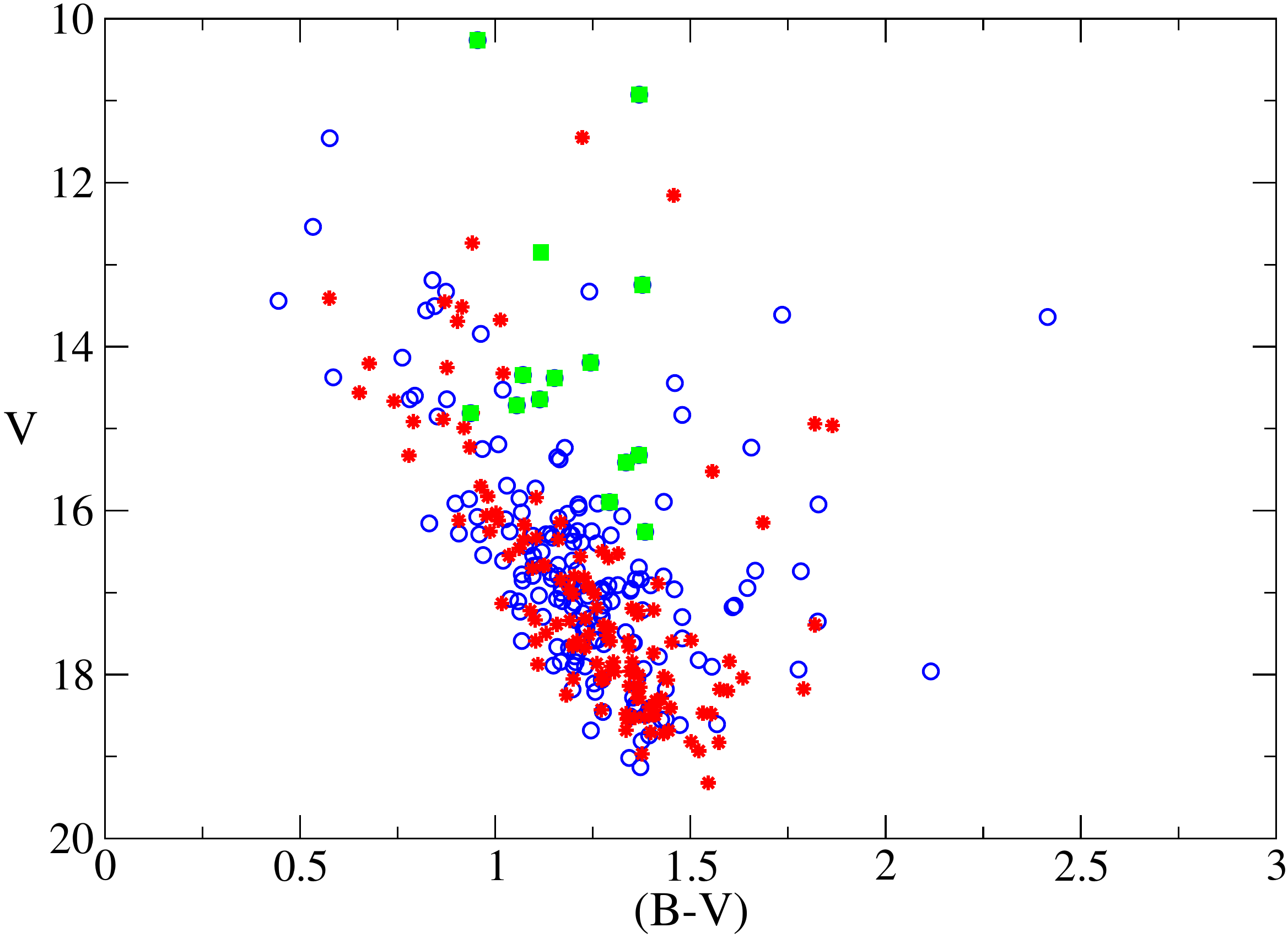}
}
\caption{\textit{Left}: $V$ magnitude against $(U-B)$ colour for the two frames observed in the region of Berkeley~90. \textit{Right}: $V$ magnitude against $(B-V)$ colour for the two frames observed in the region of Berkeley~90\label{BVtodos}. In both diagrams, blue open circles are stars observed in the field centred on the cluster; red stars are objects observed in the area to the southeast of Berkeley~90; and green filled squares are early-type stars spectroscopically observed\label{UBtodos}.}
\end{figure*}

Initially, we plot the  $V$/$(U-B)$ and $V$/$(B-V)$ diagrams for all stars in the two frames (see Fig.~\ref{UBtodos}). 
The stars belonging to the field of Berkeley~90 (open blue circles in Fig.~\ref{UBtodos}) are numbered until {\bf 182} in Table~\ref{tab:coorBe90}. The rest of the numbers in this table (red stars in Fig.~\ref{UBtodos}) are stars located in the southeast Berkeley~90 frame. The green squares are stars with spectra as well. We can see that the two samples occupy the same position in the two diagrams. 
%The first goal is to determine the physical properties of the cluster Berkeley~90: distance and age. Further, we observed two different areas (one additional field different from the cluster centre) because we wanted to know the characteristics of the Southeast population, i.e. if the stars belong to these two frames are placed at the same distance and have the same age. This method it is an easy way to verify if there has been a sequential process of stellar formation or if this event has occurred at the same time over the whole area. We wanted to obtain information about the area bordering with the cluster as well.
%In Fig.~\ref{UBtodos} we can see that 
There is a very well-defined main sequence to the left of the $V$/$(U-B)$ diagram and many objects spread out to the right of this location. The spectroscopically confirmed early-type stars in the cluster (green squares) fit very well with the left sequence. But there is clearly differential reddening and a high degree of contamination. We must thus perform a detailed analysis to select the members of the cluster.

%\begin{figure}
%\resizebox{\columnwidth}{!}{\includegraphics[clip]{/home/amparo/work/Be90/alfosc20070709/OBJECTS/ANALISIS/GRAFICAS/UBV_diagram_todos.eps}
%}
%\caption{$V$ magnitude against $(U-B)$ colour for the two frames observed in Berkeley~90. The blue open circles are stars observed in the field centered in the cluster. Red stars are objects observed in the area of the Southeast of Berkeley~90. Green solid squares are early-type stars spectroscopically observed.\label{UBVtodos}}
%\end{figure} 

\subsubsection{Determination of the $R_{5495}$ value and a preliminary E(B-V)}
\label{chorizos}

The next step is to determine whether the extinction law in the direction of the cluster is standard. To this aim, we used the latest version of the software package {\sc CHORIZOS} \citep{jesus2004}, with the stellar grid presented by \citet{jesus2013} for the 11 brightest OB main-sequence stars. {\sc CHORIZOS} fits synthetic photometry derived from the spectral energy distribution of a stellar model convolved with an extinction
law to reproduce the observed magnitudes. We used as input the $UBVRJHK_{\textrm{S}}$ photometry and the effective temperature, $T_{\textrm{eff}}$, corresponding to the spectral types (this calibration is an adapted version of \citet{martins2005} which includes the spectral subtypes used by \citet{sota2014}.  In fact, $T_{\textrm{eff}}$ was not fixed, but allowed to vary within small limits to improve the fit. The output of {\sc CHORIZOS} is the monochromatic value of $R$ for each star, and its integrated colour excess and extinction, which are displayed in Table~\ref{tabCHO}\footnote{Note that the values derived for LS\,III~$+46\degr$11 and LS\,III~$+46\degr$12 are slightly different from those in \citetalias{jesus2015a}, because here we are using our own photometry, which includes the $R$ band, while in \citetalias{jesus2015a}, archival $UBV$ photometry was used. In any case, the differences between our photometry and published photoelectric values are within the errors of the photometry}. We calculated then the individual value of $R_{V}$ for each star using the values of $A_{V}$ and $E(B-V)$. After excluding from the calculation star {\bf 47} (because its value for $R$ deviates by more than 1-$\sigma$ from the median value), we obtain a final median value of $R_{V}=3.51\pm0.07$ which we will be using in the following analysis.

\begin{table*}
	\centering
	\caption{Parameters obtained for 13 bright stars in the cluster using the {\sc CHORIZOS} code.}
	\label{tabCHO}
	\begin{tabular}{lccccccccc} 
		\hline

Name & $\chi^2_{\textrm{red}}$ & $T_{\textrm{eff}}$(K) & LC& $E(4405-5495)$ &$R_{5495}$ & $\log\,d$ (pc) &$A_V$&$E(B-V)$\\

\hline
36   &   1.325 & 34900 & 5 & $1.668\pm0.014$ & $3.334\pm0.037$ & $3.355\pm0.003$ & $5.580\pm0.025$ & $1.646\pm0.013$\\
47   &   0.707 & 22000 & 5 & $1.282\pm0.014$ & $3.890\pm0.065$ & $3.373\pm0.008$ & $5.010\pm0.045$ & $1.267\pm0.013$\\
57   &   3.650 & 32500 & 5 & $1.496\pm0.013$ & $3.468\pm0.041$ & $3.569\pm0.003$ & $5.213\pm0.025$ & $1.480\pm0.013$\\
59   &   2.600 & 22000 & 5 & $1.268\pm0.014$ & $3.491\pm0.050$ & $3.556\pm0.004$ & $4.454\pm0.026$ & $1.252\pm0.013$\\
67   &   1.617 & 22000 & 5 & $1.552\pm0.014$ & $3.321\pm0.040$ & $3.553\pm0.004$ & $5.174\pm0.027$ & $1.524\pm0.013$\\
70   &   4.050 & 25400 & 5 & $1.386\pm0.013$ & $3.439\pm0.045$ & $3.509\pm0.004$ & $4.793\pm0.026$ & $1.368\pm0.013$\\
75   &   3.750 & 23700 & 5 & $1.341\pm0.013$ & $3.381\pm0.045$ & $3.559\pm0.003$ & $4.561\pm0.024$ & $1.324\pm0.013$\\
%84   &   1.048 & 17000 & 5 & $1.445\pm0.014$ & $3.675\pm0.055$ & $3.377\pm0.008$ & $5.324\pm0.047$ & $1.417\pm0.013$\\
88   &   1.475 & 20000 & 5 & $1.129\pm0.013$ & $3.483\pm0.056$ & $3.591\pm0.004$ & $3.960\pm0.026$ & $1.115\pm0.013$\\
90   &   3.275 & 21700 & 5 & $1.603\pm0.014$ & $3.544\pm0.040$ & $3.417\pm0.003$ & $5.696\pm0.025$ & $1.574\pm0.013$\\
92   &   0.812 & 20000 & 5 & $1.575\pm0.036$ & $3.509\pm0.097$ & $3.565\pm0.007$ & $5.537\pm0.050$ & $1.545\pm0.034$\\
176  &   1.732 & 41900 & 4 & $1.217\pm0.013$ & $3.515\pm0.052$ & $3.269\pm0.003$ & $4.311\pm0.025$ & $1.212\pm0.013$\\
178  &   2.447 & 41300 & 1 & $1.611\pm0.013$ & $3.473\pm0.039$ & $3.390\pm0.004$ & $5.612\pm0.025$ & $1.592\pm0.013$\\
283  &   3.150 & 31300 & 5 & $1.423\pm0.013$ & $3.194\pm0.040$ & $3.390\pm0.003$ & $4.574\pm0.024$ & $1.407\pm0.012$\\
 \noalign{\smallskip}
\hline
\end{tabular}
\end{table*}

\subsubsection{The reddening-free $Q$ parameter, photometric spectral types and reddening}
\label{reddening}

The reddening-free $Q$ parameter allows a preliminary selection of early-type stars. The $Q$ parameter is defined as: 

\begin{equation}
\label{Qdefinition}
Q=(U-B)-\frac{E(U-B)}{E(B-V)}(B-V)
\, .
\end{equation} 

\begin{figure*}
\resizebox{18 cm}{!}{\includegraphics[angle=0]{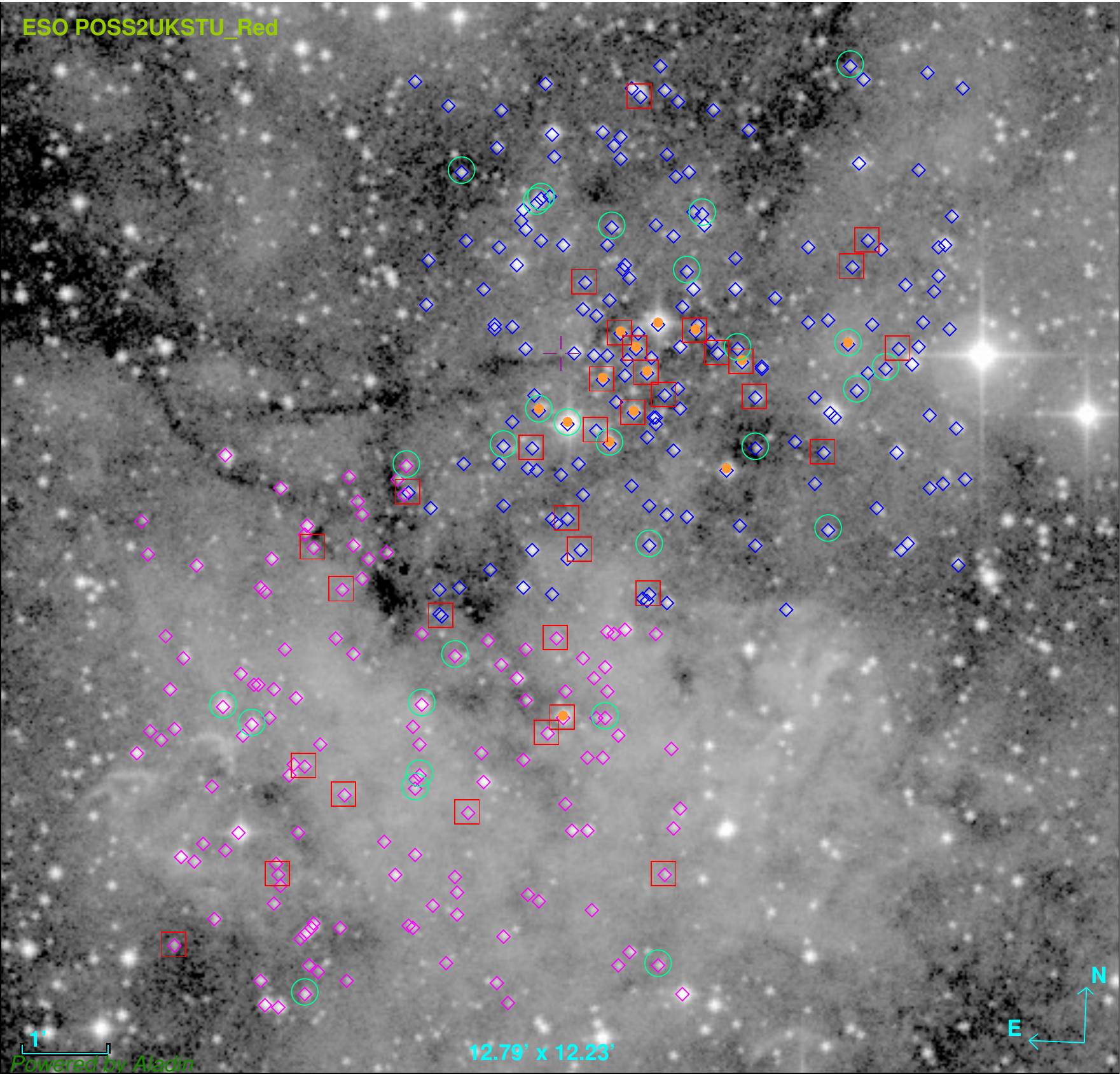}
%{reddening_map.eps}
}
\centering
\caption{Stars with optical photometry in the frame corresponding to the open cluster Be~90 (blue open diamonds) and those of the southeast frame (pink open diamonds). Open big squares in red represent stars with a value of $E(B-V)$ higher than 1.3. Open big circles in green are stars with a value of $E(B-V)$ lower or equal to 1.3. Main-sequence stars with spectra are marked with filled small circles in orange. North is up and east is left. The size of the image is $12.79\arcmin\times12.23\arcmin$. It is a DSS2 Red image.\label{mapmembers}} 
\end{figure*}

The value of $E(U-B)/E(B-V)$ is $0.72$ for a standard reddening law \citep{johnson1953}. For higher accuracy in the reddening procedure, and given that the value of $R_{V}$ is higher than standard, we calculated the $E(U-B)$ and $E(B-V)$ values for all the stars in our data set with spectra and $UBV$ photometry, using the intrinsic colours from \citet{fitzgerald1970}. We plotted the $E(U-B)/E(B-V)$ diagram for these stars and obtained the equation of the line that fits best those points using the linear least squares fitting technique. In this case, the value of the line's slope calculated is $0.81$, and we used this value when we determined the value of $Q$ for all stars in the two fields. As all B-type stars have a negative $Q$ parameter, we selected only stars with $Q\leq0$, corresponding to the early-type members of the cluster, and we assigned them photometric spectral types \citep{johnson1953}. Given our magnitude limits, we do not expect to have reached any cluster members with spectral types later than B-type. For a few stars, there is not a good match between their derived photometric spectral types and their position in the $V$/$(B-V)$ diagram. We consider these objects early-type non-members (two of them, 14 and 179, are spectroscopically confirmed in Table~\ref{tab:spTBe90M}). We identify the rest of the objects (45) as likely members of the cluster (see Figure~\ref{mapmembers}). We determined individual $E(B-V)$ values for all members, by finding their intrinsic colours. For this, we used the $(U-B)$/$(B-V)$ diagram to slide the stars along the reddening vector (with slope of 0.81) until reaching a position on the unreddened standard relation \citep{johnson1953}, from which we obtained their accurate intrinsic colour $(B-V)_{0}$. We list their values of $E(B-V)$, $(B-V)_{0}$, $V_{0}$ and photometric spectral types in Table~\ref{tab:membersBe90}.

The $E(B-V)$ values range from 1.0 until 1.6, with a median value of $1.4\pm0.2$. Most stars have $E(B-V)$ values higher than 1.2 and we find most stars between 1.3 and 1.4 (10 and 11 stars, respectively). The spatial distribution of these values is shown in Figure~\ref{mapmembers}. Open circles are members with $E(B-V)\leq1.3$ and open squares are members with $E(B-V)>1.3$. The main conclusion is that the members located near the cluster core have higher values of the reddening.

We carried out the same analysis with stars located in the southeast frame. We found a sequence of 29 B-type stars which have spectral types consistent with their positions in the $V$/$(B-V)$ diagram. They correspond to a single early-type population located at the same distance as Berkeley~90. In Table~\ref{tab:membersBe90South}, we list the values of $E(B-V)$, $(B-V)_{0}$, $V_{0}$ and photometric spectral types for all the stars considered members of this population. The $E(B-V)$ values range from 1.0 until 1.6 with a median value of $1.3\pm0.2$, consistent with that of cluster members. Most stars have reddening between 1.3 and 1.4, but the difference with the members of Be~90 is that the lower values of reddening are more populated than the higher values. In Figure~\ref{mapmembers} we can also see the distribution of these values for components of that early-type sequence, using the same symbols as in the frame corresponding to Be~90. This result confirms again the higher reddening towards the cluster core.

\begin{table}
\centering
\caption{Values of $E(B-V)$, $(B-V)_{0}$, $V_{0}$ and photometric spectral types for likely B-type members in Berkeley~90.\label{tab:membersBe90}}
\begin{tabular}{lcccc}
\hline
\hline
\noalign{\smallskip}
Star&$E(B-V)$&$(B-V)_{0}$&$V_{0}$& Photometric \\
&&&&spectral types\\
\noalign{\smallskip}
\hline
\noalign{\smallskip}
1	&	1.49	&	$-$0.02	&	13.37	&	B9\\
7	&	1.58	&	$-$0.18	&	11.37	&	B3\\
15	&	1.39	&	$-$0.18	&	11.37	&	B3\\
17	&	1.34	&	$-$0.17	&	11.50	&	B5\\
18	&	1.16	&	$-$0.13	&	12.05	&	B7\\
22	&	1.42	&	$-$0.14	&	12.00	&	B5\\
36	&	1.70	&	$-$0.32	&	7.29	&	O--B0\\
42	&	1.65	&	$-$0.17	&	11.51	&	B5\\
44	&	1.28	&	$-$0.07	&	12.86	&	B9\\
45	&	1.35	&	$-$0.19	&	11.35	&	B5\\
46	&	1.29	&	$-$0.15	&	11.75	&	B7\\
47	&	1.33	&	$-$0.26	&	9.69	&	B2\\
50	&	1.47	&	$-$0.16	&	11.73	&	B5\\
57	&	1.53	&	$-$0.29	&	8.81	&	O--B0\\
59	&	1.30	&	$-$0.25	&	10.15	&	B2\\
61	&	1.49	&	$-$0.23	&	10.69	&	B2\\
63	&	1.56	&	$-$0.19	&	11.22	&	B3\\
65	&	1.06	&	$-$0.13	&	12.13	&	B8\\
67	&	1.59	&	$-$0.25	&	9.85	&	B2\\
70	&	1.41	&	$-$0.26	&	9.43	&	B1\\
71	&	1.12	&	$-$0.10	&	12.68	&	B8\\
75	&	1.36	&	$-$0.25	&	9.86	&	B1\\
80	&	1.50	&	$-$0.21	&	11.03	&	B2\\
82	&	1.63	&	$-$0.15	&	11.84	&	B5\\
83	&	1.28	&	$-$0.11	&	12.44	&	B8\\
84	&	1.52	&	$-$0.23	&	10.55	&	B2\\
88	&	1.16	&	$-$0.22	&	10.75	&	B2\\
90	&	1.62	&	$-$0.26	&	9.63	&	B2\\
92	&	1.62	&	$-$0.24	&	10.58	&	B2\\
112	&	1.35	&	$-$0.13	&	12.16	&	B8\\
115	&	1.26	&	$-$0.17	&	12.02	&	B5\\
117	&	1.38	&	$-$0.11	&	12.44	&	B8\\
118	&	1.00	&	$-$0.09	&	12.78	&	B8\\
128	&	1.40	&	$-$0.12	&	12.25	&	B8\\
133	&	1.25	&	$-$0.13	&	12.12	&	B7\\
138	&	1.17	&	$-$0.13	&	12.16	&	B8\\
140	&	1.26	&	$-$0.25	&	10.09	&	B1\\
141	&	1.20	&	$-$0.12	&	12.23	&	B7\\
157	&	1.39	&	$-$0.20	&	11.18	&	B3\\
161	&	1.02	&	$-$0.12	&	12.34	&	B7\\
163	&	1.26	&	$-$0.11	&	12.42	&	B8\\
172	&	1.28	&	$-$0.12	&	12.30	&	B7\\
176	&	1.29	&	$-$0.33	&	5.75	&	O--B0\\
178	&	1.70	&	$-$0.33	&	4.96	&	O--B0\\
181	&	1.36	&	$-$0.12	&	12.28	&	B7\\
\noalign{\smallskip}
\hline
\end{tabular}
\end{table}

\begin{table}
\centering
\caption{Values of $E(B-V)$, $(B-V)_{0}$, $V_{0}$ and photometric spectral types for likely B-type members in Southeast Berkeley~90.\label{tab:membersBe90South}}
\begin{tabular}{lcccc}
\hline
\hline
\noalign{\smallskip}
Star&$E(B-V)$&$(B-V)_{0}$&$V_{0}$& Photometric \\
&&&&spectral types\\
\noalign{\smallskip}
\hline
\noalign{\smallskip}
185	&	1.23	&	$-$0.15	&	11.88	&	B5\\
188	&	1.32	&	$-$0.12	&	12.39	&	B7\\
191	&	0.96	&	$-$0.17	&	11.54	&	B5\\
206	&	1.39	&	$-$0.11	&	12.53	&	B8\\
210	&	1.11	&	$-$0.12	&	12.37	&	B8\\
217	&	1.12	&	$-$0.03	&	13.29	&	B9\\
219	&	1.37	&	$-$0.13	&	12.12	&	B8\\
220	&	1.29	&	$-$0.12	&	12.32	&	B8\\
222	&	1.26	&	$-$0.15	&	11.93	&	B7\\
224	&	1.37	&	$-$0.13	&	12.14	&	B7\\
232	&	1.35	&	$-$0.19	&	11.40	&	B3\\
236	&	1.27	&	$-$0.25	&	9.87	&	B1\\
237	&	1.24	&	$-$0.12	&	12.30	&	B7\\
238	&	1.25	&	$-$0.31	&	8.34	&	O--B0\\
239	&	1.18	&	$-$0.27	&	9.38	&	B1\\
250	&	1.11	&	$-$0.13	&	12.17	&	B7\\
254	&	1.12	&	$-$0.16	&	11.76	&	B5\\
255	&	1.33	&	$-$0.13	&	12.13	&	B7\\
260	&	1.38	&	$-$0.13	&	12.16	&	B7\\
265	&	1.45	&	$-$0.11	&	12.50	&	B8\\
268	&	1.14	&	$-$0.13	&	12.13	&	B7\\
269	&	1.12	&	$-$0.14	&	11.89	&	B7\\
271	&	1.38	&	$-$0.12	&	12.34	&	B8\\
278	&	1.31	&	$-$0.12	&	12.36	&	B7\\
283	&	1.44	&	$-$0.32	&	7.80	&	O--B0\\
291	&	1.62	&	$-$0.05	&	13.13	&	B9\\
295	&	1.49	&	$-$0.05	&	13.16	&	B9\\
297	&	1.65	&	$-$0.10	&	12.69	&	B8\\
301	&	1.98	&	$-$0.19	&	11.24	&	B5\\
\noalign{\smallskip}
\hline
\end{tabular}
\end{table}

\subsubsection{Determination of distance and age}
\label{distance}

\begin{figure}
\resizebox{\columnwidth}{!}{\includegraphics[clip]{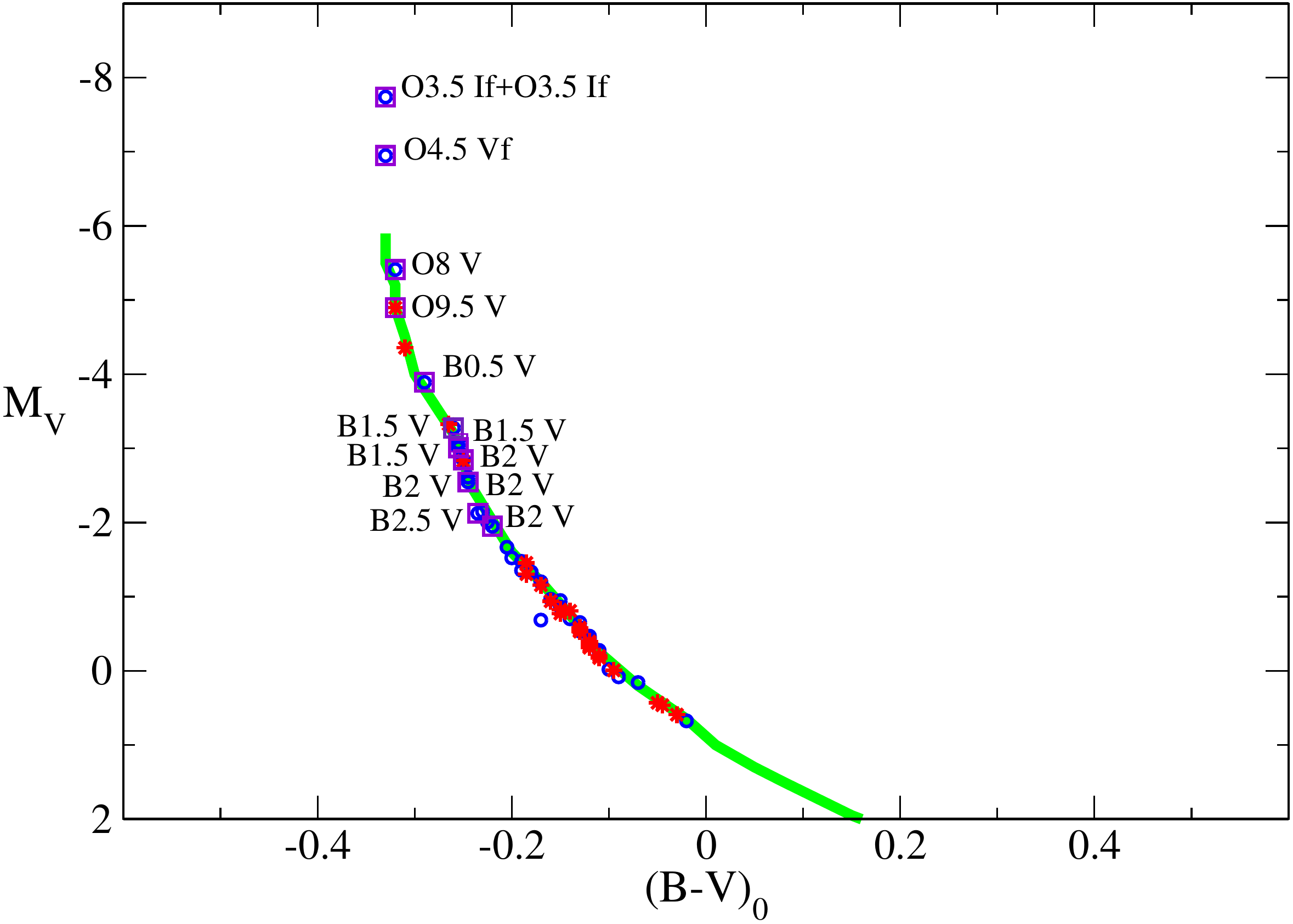}
}
\caption{Absolute magnitude $M_{V}$ against intrinsic colour $(B-V)_{0}$ for Berkeley~90 and southeast of Berkeley~90. Blue open circles are stars observed in the field centred on the cluster. Red stars are objects observed in the area to the southeast of Berkeley~90. Open violet squares are early-type stars spectroscopically observed. The solid line represents the ZAMS from \citet{kaler1982}\label{MvBV0todos}.} 
\end{figure}

We plot the $M_{V}$/$(B-V)_{0}$ diagram for all members in the fields of Berkeley and southeast of Berkeley~90 (see Fig.~\ref{MvBV0todos}). We use the observational ZAMS from \citet{kaler1982}, shifted to different distance moduli, to determine a distance modulus (DM) to the cluster of $12.7\pm0.3$, where we assume a relatively conservative error in the by-eye fit of $\pm0.3$. This value corresponds to a distance of $3.5^{+0.5}_{-0.5}$~kpc. We can test this result by comparing the absolute magnitudes obtained in this diagram for each known spectral type to the values in the spectral type calibration by \citet{turner1980}. All the B-type stars with photometric and spectroscopic spectral types match the calibration if its intrinsic dispersion (due not only to the finite size of the subtypes, but also to physical causes such as fast rotation, inclination to the line of sight and binarity) is taken into account. In contrast, O-type stars appear systematically too bright. Star \textbf{283} (O9.5\,V) is more than half a magnitude too bright, while star \textbf{36} is almost one magnitude too bright. LS\,III~$+46\degr$12 is also, as discussed at length in \citetalias{jesus2015a}, much brighter than expected. Though we cannot offer an explanation with the current data, this excess brightness very likely points towards a high multiplicity in all the O-type stars. In the case of \textbf{36}, we have to note that its H$\alpha$ photometry in the IPHAS catalogue (see figure~7 in \citetalias{jesus2015a} and the corresponding discussion, where it is referred to as 2MASS~J20350798+4649321), places it in a region typical of emission-line stars. There is no indication of such behaviour in our spectrum, while Oe stars of such an early spectral type are extremely rare in the Milky Way \citep{negueruela2004}. Finally, we note that this distance for the cluster is compatible within their respective errors with the average of the distances obtained from the {\sc CHORIZOS} analysis $d=3.0\pm0.6$~kpc (the error here is the standard deviation)\footnote{Note that {\sc CHORIZOS} makes the implicit assumption that all stars are single, a fact which easily explains the lower average distance and the high dispersion in individual values}.

Given that there are no evolved stars in the cluster (even the O3.5\,If* supergiants are in all likelihood main sequence stars; see discussion in \citetalias{jesus2015a}), we cannot fit any isochrone in the $M_{V}$/$(B-V)_{0}$ diagram to determine the age. We can only assert that the age of the cluster is less than 3~Ma, because we still have a full complement of O-type stars on the main-sequence. Stars considered members in the two frames form a single sequence, and so are located at the same distance. In the $(U-B)$/$V$ CMD (Fig.~\ref{UBtodos}) we can see that the two populations fit very well to the ZAMS and form a single sequence of early-type stars, while in the $(B-V)$/$V$ CMD stars from the Southeast frame occupy positions consistent with lower reddening. Therefore this second population has an age similar to that of the cluster. In the next section, we analyse the relationship between these two populations.

Given its Galactic coordinates ($l=84\fdg9$, $b=+3\fdg8$) and its distance ( $d=3.5^{+0.5}_{-0.5}$~kpc), Berkeley~90 is very likely to be placed in the Perseus arm, which has tracers at $d\la4$~kpc around $l=95\degree$\,--\,$100\degree$ \citep{choi2014}, as the Local arm does not seem to extend in this direction \citep{xu2013}. Unfortunately, there are no other clusters catalogued in the area that could be used to constrain structure in this Galactic region. Other clusters with about same Galactic longitude ($\sim 85\degree$) have very different latitudes. We can cite Berkeley~88, NGC~7024, NGC~6996 and NGC~6989 (with $b=+6\fdg5$, $b=-3\fdg9$, $b=+0\fdg1$ and $b=+0\fdg3$, respectively), but there are no in-depth studies on any of them providing accurate properties yet.

\subsubsection{Near-IR photometry}
\label{ir}

\begin{figure*}
\centering
\resizebox{\columnwidth}{!}{\includegraphics[clip]{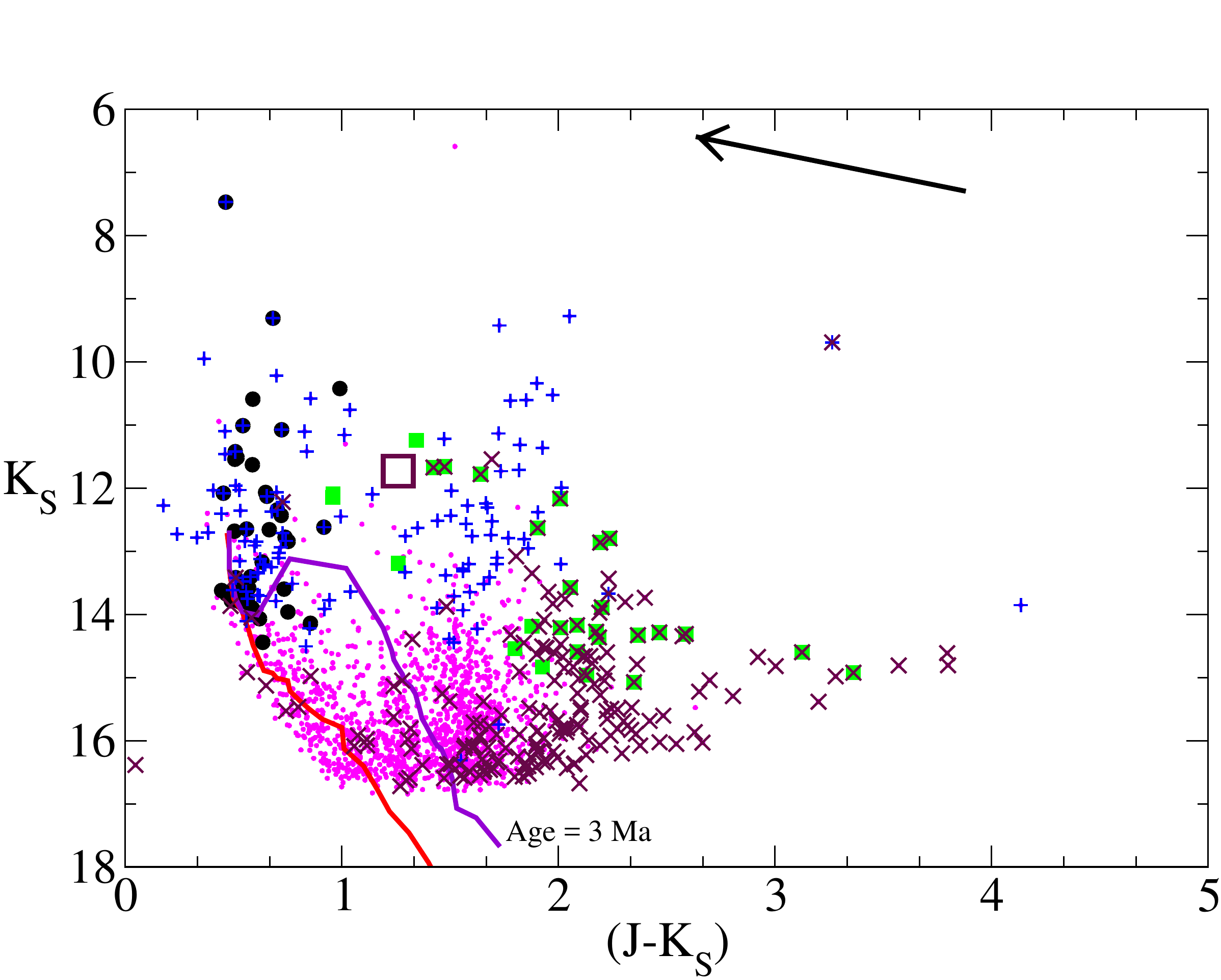}
}
\resizebox{\columnwidth}{!}{\includegraphics[clip]{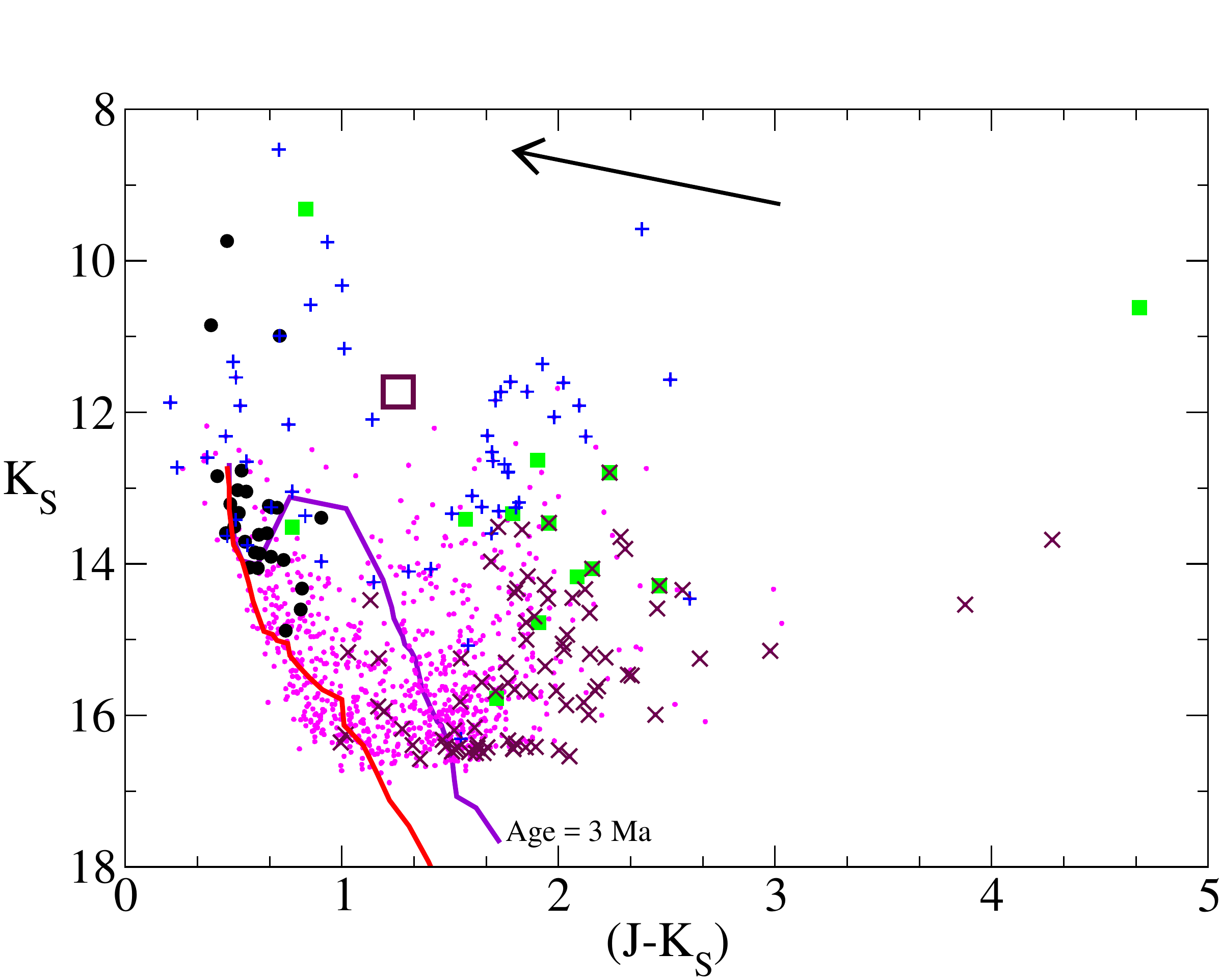}
}
\caption{\textit{Left}: $K_{\textrm{S}}$ magnitude against $(J-K_{\textrm{S}})$ colour for Berkeley~90. \textit{Right}: $K_{\textrm{S}}$ magnitude against $(J-K_{\textrm{S}})$ colour for the area of southeast Berkeley~90. We use pink dots to represent all stars with near-IR photometry. The likely B-type members (from the optical analysis) are represented as the black filled circles. Stars identified as discless from the {\textit WISE} photometry are marked with the blue pluses. PMS stars with discs according to the {\textit WISE} criteria are represented by green filled squares. Stars with $Q\le-0.1$ are the brown crosses. The big open square indicates the approximate position for red clump stars at the distance and reddening of the cluster. The red line is the early main sequence and the purple line is the 3 Ma PMS isochrone, both taken from \citet{siess2000}. The reddening vector $K_{\textrm{S}}=0.68\cdot(J-K_{\textrm{S}})$ of length 1.5 mag is indicated by the arrow. See text for further details\label{IR}.}
\end{figure*}

The centre and size of Berkeley~90 were set by \citet{tadross2008} using 2MASS data. This area, corresponding to a diameter of $5\arcmin$, was fully covered by our optical photometry (a $6\farcm5\times6\farcm5$ frame). However, 2MASS data clearly suffer from incompleteness due to confusion near the cluster core, as many of our optically identified stars lack 2MASS counterparts. Therefore we used the near-IR photometry obtained from the UKIDSS images covering the same size as our optical images. First, we plot the $K_{\textrm{S}}$/$(J-K_{\textrm{S}})$ diagram for Berkeley~90 (see Fig.~\ref{IR}). We used the values of $E(B-V)$ and distance modulus obtained in Section~\ref{reddening} and Section~\ref{distance} to draw the early-main sequence, ranging from spectral type M6 until B3, with $Z=0.02$ and no overshooting. The early main sequence is defined as the stellar locus at the time when the star settles on the main sequence after the CN cycle has reached its equilibrium, which only affects stars with a mass higher than $1.2\:$M$_{\sun}$. We plot the pre-main sequence (PMS) isochrone of 3~Ma with the same parameters as well. Both tracks are taken from \citet{siess2000}. 

We obtained a median value of $R_{V}=3.51\pm0.07$ in Section~\ref{chorizos} which deviates from the standard value of $R_{V}=3.1$ and, therefore, we cannot use the standard relationship between $E(B-V)$ and $E(J-K_{\textrm{S}})$. Using the extinction laws implemented in {\sc CHORIZOS}, we determined the values of $E(J-K_{\textrm{S}})$ and $A_{K_{\textrm{S}}}$ that correspond to $E(B-V)\sim1$, the lowest value measured for cluster members (see Table~\ref{tab:membersBe90}) when $R_{V}=3.51$ (finding $E(J-K_{\textrm{S}})= 0.65$ and $A_{K_{\textrm{S}}}=0.44$) and displaced the main-sequence and PMS isochrone to these values. In Fig.~\ref{IR}, we use pink dots to represent all stars with near-IR photometry. The likely B-type members (from the optical analysis) are represented as black filled circles. There is a population of faint sources close to the main-sequence, but a much larger population around the 3~Ma isochrone, which is the estimated maximum age for the cluster. Taking this into account, we can interpret the pink dots with $(J-K_{\textrm{S}})\le1.2$ as foreground stars, while the rest of the stars are the PMS population of the cluster together with some field contamination that we estimate below. 

To characterize the PMS stars, we carried out the same procedure as in Section~3.2.9 from \citet{marco2016}, using {\textit WISE} data and the near-IR $Q$ index. There are 146 stars with {\textit WISE} counterparts in the field of Be~90, and 69 in the field of the southeast of Be~90. We used the $(K_{\textrm{S}}-W1)$ against $(W1-W2)$ and $(H-K_{\textrm{S}})$ against $(W1-W2)$ diagrams to classify the stars as either objects with disc or discless, following the criteria described in \citet{koenig2012} and \citet{koenig2014}. We were not able to utilize diagrams based on the $W3$ or $W4$ bands, because dust emission from the molecular cloud dominates these bands for all sources in this area. To reach fainter stars, we can use the $(J-H)-(H-K_{\textrm{S}})$ diagram, or equivalently the IR $Q$ reddening-free index. Stars with discs or strong infrared excess have negative values of $Q$ \citep{negueruela2007} and we take $Q\le-0.1$ as the condition to identify such stars.

In Fig.~\ref{IR}, stars identified as discless from the {\textit WISE} photometry are marked with the blue pluses. Most of them match with B-type main-sequence stars or obvious non-members. PMS stars with disks are represented by the green filled squares. Stars with $Q\le-0.1$ are the brown crosses. These two populations represent the PMS population living in the cluster. They are placed on the right side of the diagram because they have an excess in the $(J-K_{\textrm{S}})$ colour arising from the disk. As {\textit WISE} photometry is relatively shallow, we only find systems with disks among the brightest targets (down to $K_{\textrm{S}}\approx15$~mag). The green squares with $K_{\textrm{S}}\la14$~mag likely correspond to the massive PMS stars ($M\geq3M_{\sun}$), while the brown crosses extend to fainter magnitudes, covering the same range of $(J-K_{\textrm{S}})$. We interpret them as lower-mass PMS objects. To estimate the level of contamination by background stars, we plot the approximate position for red clump stars at the distance and reddening of the cluster as a big open square. We can project this locus along the reddening vector (drawn on the top of the figure) to see their possible positions in the diagram. We do not find many objects compatible with these locations, and most of them have characteristics that identify them as PMS stars. Indeed, if we select stars with $Q\ga0.35$ (typical of red giants), we find that most of them have $K_{\textrm{S}}$ magnitudes in the range 16\,--\,17 and, therefore, cannot be red giants, because they are very far away from their possible positions. They must then be foreground red dwarfs. The low contamination by background red giants is not surprising, given the moderately large distance to the cluster and its height above the Galactic plane.     

Furthermore, we plot in Fig.~\ref{IR} the $K_{\textrm{S}}$/$(J-K_{\textrm{S}})$ diagram for the southeast of Berkeley~90 as well. We took the near-IR photometry from the UKIDSS images covering the same area as our optical frames. We use the same values of the parameters as for Berkeley~90 to build the diagram. The symbols represent the same kinds of objects as in the Berkeley~90 diagram. We can also see a PMS population living in this area, associated with the stars considered members based on the optical photometry. In fact, the distribution of sources in this CMD is very similar to that found in the CMD for the cluster.

\begin{figure*}
\resizebox{18 cm}{!}{\includegraphics[angle=0]{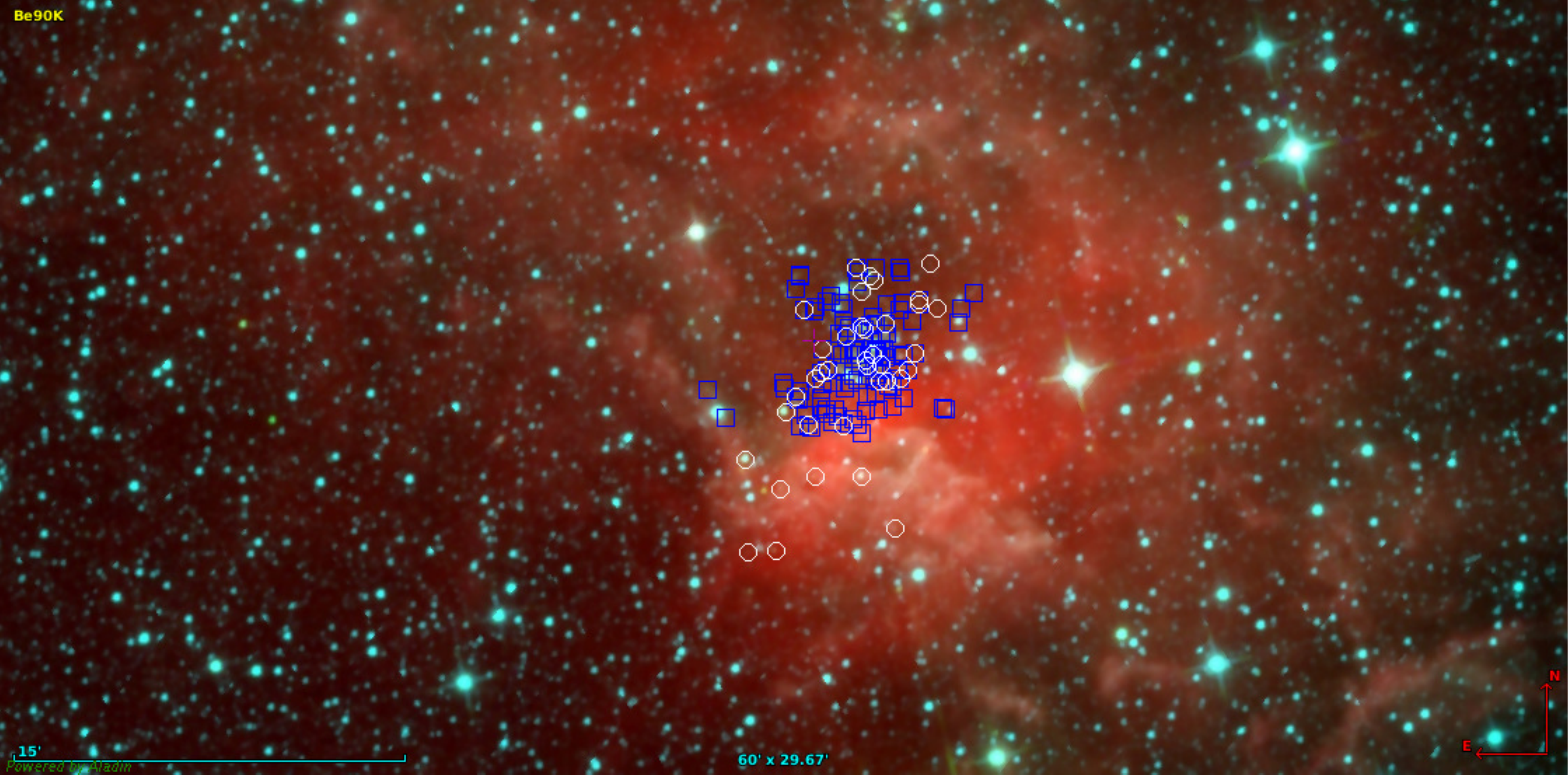}
%{reddening_map.eps}
}
\centering
\caption{Spatial distribution of the PMS stars in the two areas studied. The image, downloaded from Aladin, is provided by the {\textit WISE} mission. The filters $W1$, $W2$ and $W4$ are represented by blue, green and red, respectively. White circles represent stars with discs. Blue squares are low mass stars selected with the infrared $Q$ index and occupying the same location in the $K_{\textrm{S}}$/ $(J-K_{\textrm{S}})$ diagram that the stars with discs but reaching fainter $K_{\textrm{S}}$ magnitudes. The Herbig Be star (S238) is the easternmost star with disc, to the southeast of the cluster (clearly recognizable by the high concentration of PMS stars), on the bright dust shell. North is up and east is left. The size of the image is $1\degree\times29.67\arcmin$\label{PMS}.} 
\end{figure*}

\begin{figure}
\resizebox{\columnwidth}{!}{\includegraphics[clip]{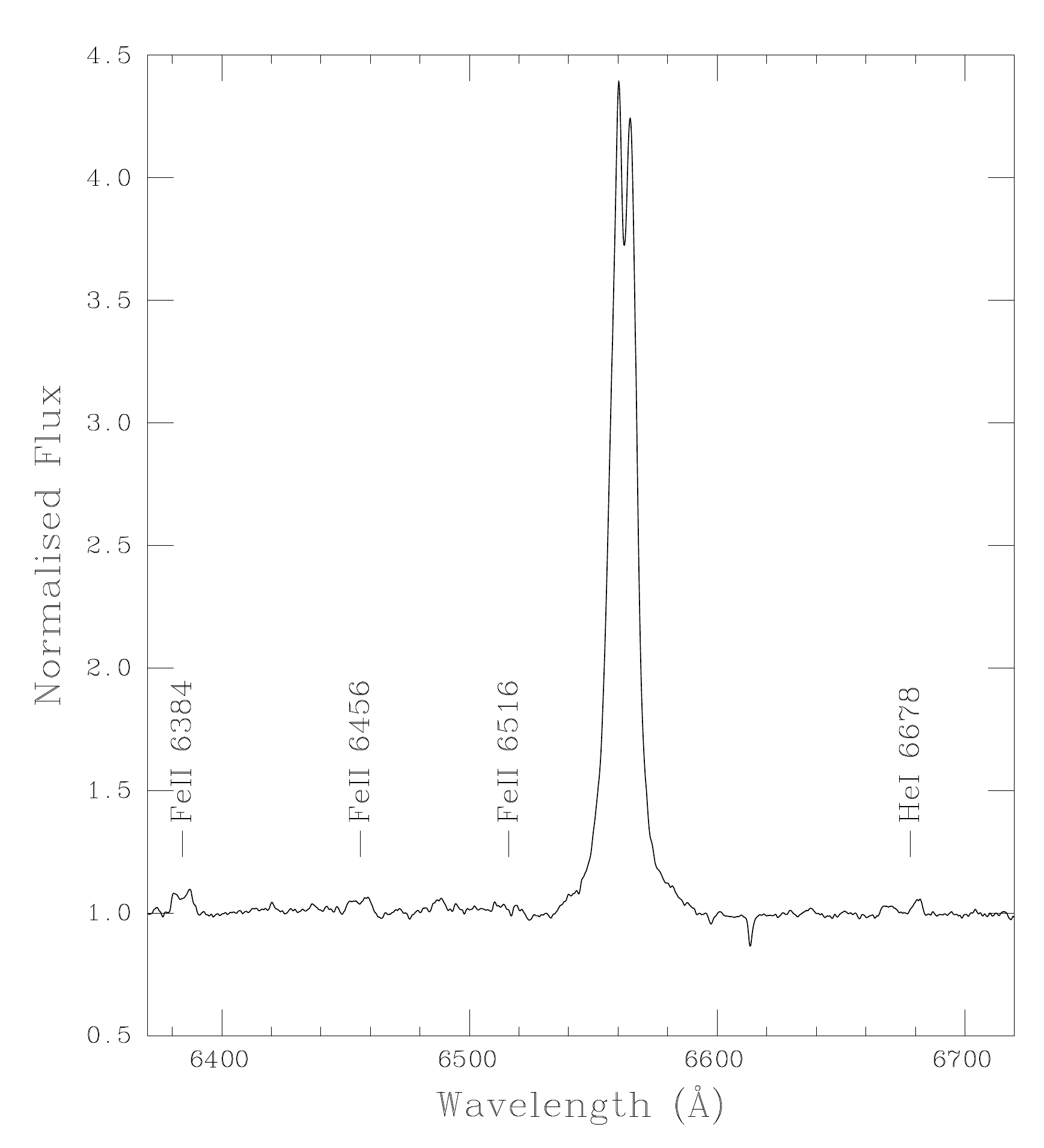}
}
\caption{Intermediate-resolution spectrum of star \textbf{238} around H$\alpha$ showing emission lines typical of an early-type Be star. The equivalent width of H$\alpha$ is $-45\,${\bf \AA}.\label{alpha}} 
\end{figure}

We represent the spatial distribution of the PMS stars in the two regions studied in Fig.~\ref{PMS}. We plot the stars classified as PMS stars with discs from {\textit WISE} data as white circles; the rest of the PMS low-mass population, surrounding these stars in the $K_{\textrm{S}}$/$(J-K_{\textrm{S}})$ diagram, as blue squares (in this map, we are using only stars with $(J-K_{\textrm{S}})\geq1.5$ and $K_{\textrm{S}}\leq16$ to avoid the noise caused by objects with low-precision photometry). We can see that both kinds of PMS objects are detected in the cluster frame, but the low-mass population is not seen in the lower part of the southeast frame. All the objects from its $K_{\textrm{S}}$/$(J-K_{\textrm{S}})$ diagram fulfilling $(J-K_{\textrm{S}})\geq1.5$ and $K_{\textrm{S}}\leq16$ (i.e. low-mass PMS candidates) are concentrated near the border line with the Be~90 frame (to the North). The {\textit WISE} image shows that the area where we do not detect low-mass PMS stars has a much higher amount of thick dust that appears projected in front of the population. This suggests two possible (non-exclusive) explanations for the detection of only {\textit WISE} candidates to intermediate-mass PMS stars in this part of the frame. On the one hand, the population may still be embedded, with only the brightest, most massive PMS stars being detectable. On the other hand, the population in this area could be younger than the cluster, with the low-mass stars still being formed. Since we are not able to determine the age of the cluster, but an age of 2\,--\,3~Ma is probable \citepalias[see also the discussion in][]{jesus2015a}, there is some room for an age spread, implying sequential, perhaps triggered, star formation to the South of the cluster. Some support for this hypothesis comes from the detection in this area of a very massive PMS star. This object, which can be seen in Fig.~\ref{IR} (right-hand panel) as the  brightest early-type star in $K_{\textrm{S}}$, is star \textbf{238}, whose colours identify it as a very early star. A red spectrum is displayed in Fig.~\ref{alpha}, showing H$\alpha$ and \ion{He}{i}~6678\,\AA\ in emission. Its blue spectrum shows that it is a Herbig Be star with spectral type between B0.5 and B1. This is necessarily a very young object.

In view of this, we can speculate that star formation is still occurring close to the bright illuminated rims to the south of our field, where at least one ``elephant-trunk'' structure can be seen in the DSS2 images close to star \textbf{238} (see also Fig.~\ref{PMS} where \textbf{238} is the easternmost star with disc). There is no measurable age difference between the core of the cluster and the area immediately south of it (i.e. the top half of our southwest frame), but there does seem to be a difference with respect to the region rich in gas and dust, as only massive PMS stars can be seen in the mid-IR in this region. So, we conclude that there is some evidence for sequential star formation in this area. The population to the south of the cluster has the typical characteristics generally associated with triggered star formation (even though we have no obvious evidence for triggering here beyond the age difference): a smaller population (we observe the presence of less main-sequence and pre-main-sequence stars) and an absence of the most massive stars (there is only an O9.5\,V star in the region immediately south, and only early-B stars in the region still rich in dust, against the early-O stars in the cluster).

\section{Discussion}
\label{discussion}

\subsection{Cluster mass}

One of the most important issues in modern astrophysics is the determination of the IMF with which a group of stars is formed. We carry out this work in the open cluster Be~90. We have determined the upper-main-sequence population with accuracy using photometry and spectroscopy. We know how many stars there are between the different spectral ranges from O-type until B9-type stars. We can assign the stellar masses for each one using the calibrations from \citet{harmanec1988} and \citet{martins2005} (see Table~\ref{mass}). The masses of the two components of LS\,III~$+46\degr$11 are not known. The orbital solution only provides minimum masses of 39 and $36\:$M$_{\sun}$ if we were seeing the system exactly edge on. We will assume conservative values of $50\:$M$_{\sun}$, even though some studies suggest higher masses for this spectral type \citep[e.g.][]{massey05}.

\begin{table}
\centering
\caption{Values of spectral types for early-type members and their mass from \citet{harmanec1988} and \citet{martins2005}, number of stars in the cluster, and total mass for each spectral type in Berkeley~90.\label{mass}}
\begin{tabular}{lccc}
\hline
\hline
\noalign{\smallskip}
Spectral type& Mass ($M_{\sun}$)&Number&Total mass ($M_{\sun}$)\\
\noalign{\smallskip}
\hline
\noalign{\smallskip}
O3.5\,If$^{*}$&50&2& 100\\
O4.5\,V(f)&42&1& 42\\
O8.5\,V&20.8&1& 21\\
B0\,V&14.6&0&0\\
B0.5\,V&13.2&1&13\\
B1\,V&11.0&1&11\\
B1.5\,V&10.0&3&30\\
B2\,V&8.6&7&63\\
B2.5\,V&7.3&1&7\\
B3\,V&6.07&4&24\\
B5\,V&4.36&7&28\\
B7\,V&3.38&7&21\\
B8\,V&2.91&9&27\\
B9\,V&2.52&2&6\\
\noalign{\smallskip}
\hline
\end{tabular}
\end{table}

Given our magnitude limit, the cluster parameters, and the range of $E(B-V)$ found in the cluster, we can only be sure that we are complete until spectral type B7. The IMF for stars more massive than the Sun in the Solar neighbourhood was determined by \citet{salpeter1955}:
\begin{equation}
\xi(M)=\xi_{0}M^{-2.35}
\end{equation}  
where $\xi_{0}$ is the constant which sets the local stellar density. 
In our case, taking the Salpeter law, we can determine the number of stars ($N$) that form with masses between $M_{1}$ and $M_{2}$, by integrating the IMF between these limits: 

\begin{equation}
%\begin{align*}
N=\int_{M_{1}}^{M_{2}} \xi(M) dM = \xi_{0}\int_{M_{1}}^{M_{2}} M^{-2.35} dM \\ 
=\frac{\xi_{0}}{1.35}[M_{1}^{-1.35} - M_{2}^{-1.35}]
%\end{align*}
\end{equation}

First, we needed to estimate the value of $\xi_{0}$. For that, initially we chose a box from B2 until B7 (from $9\:$M$_{\sun}$ until $3\:$M$_{\sun}$) where there are 26 stars and we are certain of completeness. We obtained a value of $\xi_{0}\approx200$. Choosing a box between B0 and B2, where there are 12 stars and we are complete as well, the value that we get for $\xi_{0}$ is about 470. We have two different values for the same constant, using the distribution of the population of Be~90. Clearly, if we use these two values of $\xi_{0}$ to calculate the number of stars between $50\:$M$_{\sun}$ and $1\:$M$_{\sun}$, we will obtain two different results. In principle, we should have the same value for the constant, using the number of stars from whatever size of box where we are counting all stars belonging to the box.  
  The obvious conclusion is that, with such poor sampling of the upper-IMF, the size of the box one chooses may determine the results. Statistical fluctuations in the upper-main-sequence may affect the calculation of $\xi_{0}$. 

With this information in mind, we decided to estimate the total mass of the cluster using the value of the constant $\xi_{0}$ provided by the box covering the [B2-B7] range, because it contains more stars and the mass range is better constrained. The total mass in stars born with mass $M_{1}=50\:$M$_{\sun}\geq M \geq M_{2} =1\:$M$_{\sun}$ should be:

\begin{equation}
%\begin{align*}
M=\int_{1 M_{\sun}}^{50 M_{\sun}} M \xi(M) dM \\
=\xi_{0}\int_{M_{1}}^{M_{2}} M^{-1.35} dM = 427 M_{\sun}\\
%\end{align*}
\end{equation} 

This result is the mass of the cluster down to solar-type stars. To estimate the low-mass component, we can use the \citet{kroupa2003} standard IMF, for which the total mass of the cluster should be around twice this value. This means that Berkeley~90 is an intermediate-mass cluster with a total mass close to, but below, 1000$\:$M$_{\sun}$ which still contains three very massive stars in the upper-main sequence. We can see in Table~\ref{mass} that the sum of the total mass in early-type stars is 393$\:$M$_{\sun}$. But, if we ignore these three most massive stars (contributing around 150$\:$M$_{\sun}$), we are left with only $240\:$M$_{\sun}$, which is more compatible with the standard IMF and the estimated total mass of the cluster (about one third of the total mass). Again, we see how our calculations are dominated by very small number statistics and thus trying to calculate an observational slope for the IMF in Berkeley~90 would not make sense. Moreover, this estimate does not take into account the effect of binarity, which would take the mass of the cluster slightly above $1\,000\:$M$_{\sun}$.

In figure~1 from \citet{weidner2010} we can see different relationships between the mass cluster and the mass of the most massive star in the cluster. In this case, the value for Be~90 ($\log\,M_{\textrm{cl}}\la3$, $\log\,M_{\textrm{max}}\ga1.7$) is above all relations, but not far from the values given by \citet[short-dashed-long-dashed line in their figure]{oey2005}  or  \citet[short-dashed line]{elmegreen2000}. It may be argued that, in order to have a fair comparison with other clusters in the literature, the population to the southeast of Be~90 should also be taken into account. However, given the number of stars detected, this region could add at most 50\% to the total mass, still leaving the total population of the region around $1\,500\:$M$_{\sun}$, and not changing the conclusions in any substantial way.

\subsection{A search for runaway stars}

A high fraction of the O-type field stars (22/43) can be considered to be runaway star candidates, based on their present or former peculiar space velocity, and the vicinity of some of them to very young clusters ($\le 10$ Ma) \citep{dewit2005}. This is expected, because $N$-body simulations of massive clusters predict that the fraction of stars expelled by dynamical processes from the clusters may be as high as (20-25)\% of O-type stars and (10-15)\% of early-B type stars \citep{fpz11,oh2016}. 

Be~90 is not considered a massive cluster in the Milky Way because its mass is very far away from the minimum value of $10\,000\:$M$_{\sun}$. However, it is not only the home of the one of the most massive systems in the Galaxy (LS\,III~$+46\degr$11), but also contains at least one other early-O massive star. In contrast, its population of late-O and early-B (B0\,--\,B2) stars is scarce (only 12 stars). This distribution of stars is unusual. There are other similar cases \citep[e.g.][]{negueruela2008}, but it is not very common. Even though the ejection of $\sim20$\% of the massive stars would not seriously affect the conclusions with respect to the relation between the cluster mass and the most massive member, it would make the observed IMF still more top-heavy, favouring even more the idea of random sampling. 

We know that the highest possible velocity of dynamical ejection is $\sim100\:\textrm{km}\,\textrm{s}^{-1}$, with values of 10\,--\,$20\:\textrm{km}\,\textrm{s}^{-1}$ being more typical of this scenario. In the extreme case, a runaway star could travel 1~pc in 10\,000 yr, while for a typical case, 50\,000~yr would be required. Be~90 is located at a distance of $3.5^{+0.5}_{-0.5}$~kpc and its maximum possible age is about 3~Ma. So, considering the upper limit on age of 3~Ma, the distance travelled is 300~pc for the first case and 30~pc for the typical case. At the distance of the cluster, 300~pc are equivalent to $\sim 5\degree$, while 30~pc are $\sim 0.5\degree$. While any search over the larger radius would be impractical, the smaller distance is comparable to the apparent size of  Sh2-115 in the sky. We have searched for candidate runaway O-type stars in
this region, using archival data and the techniques outlined in \citet{negueruela2007}.

Initially, we selected all stars in 2MASS with good quality flags and photometric errors $<0.04$~mag in all bands and then plotted the $K_{\textrm{S}}$/$Q_{\textrm{IR}}$ diagram. Early-type stars form a well-defined vertical sequence between $Q\approx-0.05$ and~$0.05$. We selected stars in this range with $K_{\textrm{S}}$ magnitudes brighter than 11.0, i.e. fainter than any O-type star in Be~90. The vast majority of these objects have $(J-K_{\textrm{S}})<0.2$, indicative of very low reddening and are obviously foreground to the cluster. Many of them are already catalogued as late-B stars, such as HD~196832 or BD~$+46\degr$2968. An interesting case is BD~$+46\degr$2978, but \citet{crampton1974} classify it as B0\,III (i.e. an evolved star) and place it closer to us than the cluster. The same authors classify BD~$+46\degr$2972 as O9.5\,V placing it at a distance of 3.4~kpc. Even though these parameters suggest a connection to  Sh2-115, inspection of the 2MASS images shows that this is really a multiple system with at least two early-type stars, ruling it out as a candidate runaway.

We find a small number of objects with $(J-K_{\textrm{S}})>2$ and $K_{\textrm{S}}$ brighter than any O-type star in the cluster. These objects are in all likelihood cool stars with dusty envelopes, whose S.E.D. reproduces that of reddened early-type stars \citep{comeron2002}. Indeed, one of them is a catalogued Carbon star, CGCS~4916. We confirmed this possibility by checking their {\textit WISE} colours. In all cases $(W1-W2)$ is positive and much higher than the value $\approx0.0$ expected for early-type stars. We are finally left with a small number of candidates at low $(J-K_{\textrm{S}})\approx0.3$ and four further candidates with $(J-K_{\textrm{S}})$ slightly higher than cluster members. We used
the APASS catalogue to obtain the $B$ and $V$ magnitudes for these objects. A few of them, such as TYC 3573-1518-1, are so bright ($V\sim10$) that they should be listed in the LS catalogue if they were OB stars. In addition, for example,  TYC 3573-2322-1 has \textit{Tycho} proper motions, and its projected motion is towards the cluster. In the end, there are only five stars that we cannot rule out as OB stars \textit{and} candidate runaways, and we list them in Table~\ref{candis}\footnote{This process also shows that there is no sizeable population of early-type or emission-line objects around the cluster except in area covered by our southeast frame.}. Given their magnitudes and colours, none of them can be earlier than an O9\,--\,B0 star at the distance of the cluster. We find no candidate for an earlier-type runaway. This suggests that the number of ejections has been very low, as expected for the low mass of the cluster.

\begin{table*}
\centering
\caption{Stars within $30\arcmin$ of Be~90 whose colours and magnitudes are compatible with OB stars at the cluster distance. $B$ and $V$ magnitudes are from APASS, while infrared magnitudes are from 2MASS.\label{candis}}
\begin{tabular}{lcccccc}
\hline
\hline
\noalign{\smallskip}
2MASS & Other & $B$ & $V$ & $J$ & $H$ & $K_{\textrm{S}}$\\
\noalign{\smallskip}
\hline
\hline
\noalign{\smallskip}
20351892+4632438   & TYC 3573-1354-1 &   12.5  &  11.8  & 10.40 & 10.17 & 10.06\\
20364196+4642428   & TYC 3573-1558-1 &   12.1  &   11.4  &  10.07 &  9.85 & 9.74\\
20355132+4700204   & $-$ &            15.8 &   14.2 &   10.61  &  10.15  &   9.89\\
20365851+4710569   & $-$  &          15.2 &   13.8 &   10.73  &  10.29  & 10.08\\
20343597+4719192   & $-$   &         14.6 &   13.5 &   10.66  &  10.25  & 10.02\\
\noalign{\smallskip}
\hline
\end{tabular}
\end{table*}

\subsection{The environment of O3-type stars in the Galaxy} 

The most massive stars are found almost exclusively in the most massive
very young clusters, such as R136, the nuclear cluster of 30 Dor in the LMC. While the highest stellar masses measured correspond to WNh stars, such as NGC~3603 A1 \citep{schnurr08}, O2\,--\,O3 stars represent the most massive stars with absorption-line spectra \citep{walborn02}, with masses above $40\:$M$_{\sun}$ \citep{massey05}. In general, these objects are found only in the most massive Galactic clusters and associations, such as NGC 3603, Cyg OB2 or the Carina Nebula.

In the Galactic O star catalogue\footnote{{\tt http://ssg.iaa.es/en/content/galactic-o-star-catalog} \citep{jesus2013}}, we have found information about all catalogued O-type stars in the Milky Way earlier than O4 and the environments where they are living. We have a number of $\sim20$ such stars, with many being binary or multiple systems. The places where we find this type of stars are as follows:

\begin{itemize}
\item The Carinae  Nebula  (NGC  3372)  is  the  largest  nebula  in
the  southern  sky. More  than  60  O-type  stars  and  several  young
open  clusters  are  located  in  or  near  the  nebula,  including
Trumpler~14 and Trumpler~16 in the bright part of the nebula. These clusters contain a number of O3.5\,V stars, testimony to their extreme youth 
\citep{hur2012}.

\item NGC~3603 is the closest known giant \ion{H}{ii} region. Its dense stellar core is populated by a large sample of O-type stars, including some very massive WNh stars and a number of O2\,--\,3 stars \citep{melena2008}.

\item NGC~6357 is a complex composed of giant molecular clouds, \ion{H}{ii} regions and open clusters interacting with the parental gas, bubble-like structures and pillars, located around $l\sim353\degree$, at a distance of 1.7~kpc. The young open cluster Pismis~24 is the most prominent of the stellar groups in the complex. Its population is formed by a number of coeval O-type stars, with the two brightest displaying spectral O3.5\,I(f) and O3.5\,III(f). \citet{massi2015} estimate a cluster mass for Pismis~24 between $2\,000$M$_{\sun}$ and $6\,000$M$_{\sun}$, and conclude that there are indications of an early ejection of massive stars from the core of Pismis~24, suggesting a dynamically unstable environment after gas expulsion.

\item The dispersed but very massive OB associations Cygnus OB2 contains two O3\,If$^{*}$ supergiants. These two stars appear younger and more massive than the bulk of the association, which presents a main-sequence turn-off at O6\,V \citep{negal08}.

\item Finally, two stars belong to small clusters: the ionizing star of Sh 2-158 in NGC~7538 \citep{jesus2016} and HD~150\,136~AaAb, the brightest components of the central star in NGC~6193. Two other stars lie within star forming regions, but not in clusters: HD~64568, classified as O3\,V((f*))z and lying in relative isolation in the vicinity of the  moderate-size star-forming region NGC~2467, far away from any of the open clusters catalogued in this region; and the primary in Bajamar star, O3.5\,III((f*)), the ionizing star of the North America Nebula \citep{jesus2016}.  

\end{itemize}

We can therefore see that about three quarters of stars earlier than O4 are associated with massive young clusters, while the remaining objects are part of intermediate mass clusters or star-forming regions. In the case of Be~90 we have one certain and one possible early-O binary systems without a large massive population. These cases of intermediate mass population provide some circumstantial evidence for random sampling of the IMF.

\section{Conclusions}
We have carried out a comprehensive photometric and spectroscopic study of the very young open cluster Berkeley~90, host to one of the most massive stellar systems in the Milky Way, LS\,III $+46\degr$11. Be~90 is a small, compact cluster (with a radius slightly above $2\arcmin$) that lies within the \ion{H}{ii} region Sh2-115, occupying a small part of a cavity that is almost $30\arcmin$ across. Exploration of the cavity with 2MASS data reveals that, outside the cluster, there is only a sizeable young population immediately to the southeast of the cluster, towards a region characterized by bright illuminated rims and an ''elephant trunk'' structure. We have presented optical, near and mid-IR photometry for a frame centred on the cluster and a partially overlapping frame that covers this region to the southeast. 

We detect a sequence of 45 early-type stars in the cluster area with variable colour excess, between $E(B-V)=1.0$ and~1.6, with the higher values concentrating towards the cluster core. In the region to the southeast, we detect a similar sequence with very similar parameters, but a preference for the lower values of $E(B-V)$, comprising 29 stars, most of them of late-B spectral type (more easily reachable because of the lower reddening). From the analysis of the SED of stars with spectroscopy, we find a value of $R_{V}=3.5$, not unexpected in this region of the sky. Fitting of the ZAMS results in a distance of $3.5\pm0.5$~kpc. The age of the cluster is not measurable, but different arguments suggest it is between 2 and~$3\:$Ma.

Apart from the two catalogued early-O systems, we only find two other O-type stars in the region, an O8\,V in the halo of the cluster and a O9.5\,V towards the south. Estimation of a mass function is hindered by the very low number of objects in the upper bins and our incompleteness beyond spectral type B7, at least in the central area. We estimate a lower limit on the mass of the cluster of about $800\:$M$_{\sun}$ for the standard IMF. Inclusion of the population to the southeast and a rough correction for binarity could put the total mass up to about $1\,500\:$M$_{\sun}$. This is a low mass for a cluster containing stars of $\ga50\:$M$_{\sun}$, suggesting that random sampling and statistical fluctuations are very important at determining the mass of the most massive star in (at least, some) clusters.

Analysis of the near- and mid-IR photometry reveals a population of PMS stars associated with the cluster and immediately to its south. Further to the southeast, where there are signs of ongoing star formation, we identify a very early Herbig Be star, and a population of massive PMS stars in the vicinity of the gas and dust dark clouds. This represents evidence for sequential star formation, perhaps triggered by the very massive stars in the cluster.

\section*{Acknowledgements}

We thank the referee for corrections and helpful suggestions.

We thank Jes\'us Ma\'{i}z Apell\'aniz for the use of the {\sc CHORIZOS} code and comments of the manuscript. 

This research is partially supported by the Spanish Government Ministerio de Econom\'{i}a y Competitivad (MINECO/FEDER) under grant AYA2015-68012-C2-2-P. AM acknowledges support from the Ministerio de Educaci\'on, Cultura y Deporte through the grant PRX15/00030. 

This paper is based on observations made with the Nordic Optical Telescope, operated
on the island of La Palma jointly by Denmark, Finland, Iceland,
Norway, and Sweden, in the Spanish Observatorio del Roque de los
Muchachos of the Instituto de Astrof\'{\i}sica de Canarias.  

Some of the data presented here have been taken using ALFOSC, which is owned by the Instituto de Astrofisica de Andalucia (IAA) and operated at the Nordic Optical Telescope under agreement between IAA and the NBIfAFG of the Astronomical Observatory of Copenhagen.

This paper uses UKIDSS images. The UKIDSS project is defined in \citet{lawrence2007}. UKIDSS uses the UKIRT Wide Field Camera (WFCAM; \citep{casali2007}). The photometric system is described in \citet{hewett2006}, and the calibration is described in \citet{hodgkin2009}. The pipeline processing and science archive are described in \citet{hambly2008}.

{\sc iraf} is distributed by the National Optical Astronomy Observatories, which are operated by the Association of Universities for Research in Astronomy, Inc., under cooperative agreement with the National Science Foundation

This research has made use of the Simbad data base, operated at CDS,
Strasbourg (France) and of the WEBDA data base, operated at the Department of Theoretical Physics and Astrophysics of the Masaryk University. This publication
makes use of data products from  
the Two Micron All Sky Survey, which is a joint project of the University of
Massachusetts and the Infrared Processing and Analysis
Center/California Institute of Technology, funded by the National
Aeronautics and Space Administration and the National Science
Foundation.

This research was made possible through the use of the AAVSO Photometric All-Sky Survey (APASS), funded by the Robert Martin Ayers Sciences Fund.
%%%%%%%%%%%%%%%%%%%%%%%%%%%%%%%%%%%%%%%%%%%%%%%%%%



%\bibliographystyle{mnras}


%%%%%%%%%%%%%%%%%%%%%%%%%%%%%%%%%%%%%%%%%%%%%%%%%%

%%%%%%%%%%%%%%%%% APPENDICES %%%%%%%%%%%%%%%%%%%%%

\appendix

\section{Photometric Tables}

%\longtab{}{
\begin{table*}
\centering
\caption{Coordinates in J2000 and UBVR photometry for all stars in the field of Berkeley~90. Material on-line\label{tab:coorBe90}}
\begin{tabular}{lllrrrrrrrrr}
\hline
RA(J2000)& DEC(J2000)& Star& $V$&$\sigma_{V}$&$(B-V)$&$\sigma_{B-V}$&$(U-B)$&$\sigma_{(U-B)}$&$(V-R)$&$\sigma_{(V-R)}$&N\\
\hline\hline
20:35:26.97& $+$46:47:52.3& 1  & 18.615& 0.021 & 1.473& 0.030 &  1.042 & 0.062  & 0.958 & 0.029&1\\ 
20:35:27.14& $+$46:47:53.7& 2  & 19.130 & 0.033 & 1.372& 0.042&  0.614 & 0.061  & 0.941 & 0.046&1\\  
20:35:03.96& $+$46:47:57.2& 3  & 16.796& 0.024 & 1.096& 0.026&  0.463 & 0.014  & 0.655 & 0.029&3\\  
20:35:11.87& $+$46:48:02.0& 4  & 15.349& 0.019 & 1.159& 0.020 &  0.726 & 0.009  & 0.655 & 0.022&6\\ 
20:35:13.26& $+$46:48:03.1& 5  & 15.192& 0.017 & 1.008& 0.018&  0.390  & 0.009  & 0.610  & 0.020&6\\ 
20:35:13.58& $+$46:48:03.9& 6  & 17.931& 0.014 & 1.380 & 0.019&  0.632 & 0.030   & 0.837 & 0.023&3\\ 
20:35:13.06& $+$46:48:07.0& 7  & 16.912& 0.018 & 1.398& 0.020 &   0.526&  0.017 &  0.903& 0.022&6\\  
20:35:19.58& $+$46:48:07.2& 8  & 17.613& 0.012 & 1.352& 0.015&   0.759&  0.023 &  0.865& 0.019&4\\  
20:35:27.09& $+$46:48:10.3& 9  & 18.812& 0.024 & 1.375& 0.030 &   0.754&  0.058 &  0.933& 0.035&1\\  
\hline
\end{tabular}
\end{table*}
% }

%\longtab{}{
\begin{table*}
\centering
\caption{Coordinates in J2000 and near-IR photometry for stars in the field of Berkeley~90. Material on-line\label{tab:ir}}
\begin{tabular}{lllrrrrrr}
\hline
RA(J2000)& DEC(J2000)&$J$&$\sigma_{J}$&$H$&$\sigma_{H}$&$K_{S}$&$\sigma_{K_{S}}$\\
\hline\hline
20:35:10.51&   $+$46:48:55.3 &  17.527 &  0.037  & 15.983  &  0.020 &  15.560&   0.034\\
20:35:12.30&   $+$46:52:16.3 &  17.961 &  0.045  & 16.570  &  0.024 &  16.230&   0.047\\
20:35:12.91&   $+$46:50:31.2 &  18.208 &  0.043  & 16.759  &  0.025 &  16.362&   0.052\\
20:35:28.20&   $+$46:55:50.9 &  17.757 &  0.026  & 16.801  &  0.033 &  16.622&   0.039\\
20:35:09.12&   $+$46:50:43.6 &  16.722 &  0.049  & 15.962  &  0.054 &  15.875&   0.053\\
20:35:31.02&   $+$46:54:47.7 &  16.739 &  0.037  & 16.023  &  0.027 &  15.951&   0.036\\
20:35:40.14&   $+$46:53:46.5 &  17.392 &  0.048  & 16.427  &  0.025 &  16.208&   0.031\\
20:35:17.64&   $+$46:49:38.3 &  17.493 &  0.046  & 16.115  &  0.022 &  15.666&   0.031\\
20:35:43.75&   $+$46:55:53.4 &  17.295 &  0.043  & 16.491  &  0.037 &  16.358&   0.056\\
20:35:05.22&   $+$46:48:14.3 &  17.625 &  0.039  & 16.762  &  0.025 &  16.593&   0.042\\
20:35:45.18&   $+$46:55:23.3 &  16.822 &  0.024  & 16.019  &  0.037 &  15.871&   0.029\\
20:35:42.63&   $+$46:49:54.1 &  17.739 &  0.049  & 16.846  &  0.041 &  16.646&   0.050\\
20:35:38.41&   $+$46:54:46.5 &  17.228 &  0.040  & 16.450  &  0.030 &  16.313&   0.042\\
20:35:49.67&   $+$46:55:13.8 &  17.902 &  0.039  & 16.802  &  0.039 &  16.484&   0.057\\
20:35:15.89&   $+$46:52:06.8 &  16.903 &  0.033  & 16.139  &  0.027 &  16.004&   0.038\\
20:35:56.08&   $+$46:55:50.4 &  16.503 &  0.043  & 15.736  &  0.030 &  15.595&   0.039\\
20:35:27.33&   $+$46:49:15.4 &  17.461 &  0.038  & 16.659  &  0.024 &  16.496&   0.046\\
20:35:57.90&   $+$46:55:55.2 &  17.122 &  0.035  & 16.386  &  0.048 &  16.257&   0.038\\
20:35:02.95&   $+$46:48:14.2 &  17.292 &  0.035  & 16.435  &  0.020 &  16.237&   0.042\\
20:35:32.23&   $+$46:55:55.4 &  17.115 &  0.026  & 16.365  &  0.022 &  16.225&   0.038\\
20:35:15.87&   $+$46:55:39.7 &  17.517 &  0.037  & 16.681  &  0.032 &  16.490&   0.028\\	       
\hline
\end{tabular}
\end{table*}
% }

%%%%%%%%%%%%%%%%%%%%%%%%%%%%%%%%%%%%%%%%%%%%%%%%%%

% Don't change these lines
\bsp	% typesetting comment
\label{lastpage}
\end{document}